\documentclass[final,3p,times]{elsarticle}

\usepackage{amssymb}
\usepackage{mathrsfs}
\usepackage{amsfonts}
\usepackage{amsmath,amssymb,amsfonts}
\usepackage{slashed}
\usepackage{array}
\usepackage{float}
\usepackage{verbatim}
\usepackage{epsfig}
\usepackage{graphicx}
\usepackage{color}
\usepackage[colorlinks=true, linkcolor=blue, citecolor=blue, urlcolor=blue]{hyperref}
\usepackage[dvipsnames]{xcolor}
\usepackage{multirow}
\usepackage{lipsum}
\usepackage{cleveref}
\usepackage{comment}
\usepackage{physics}
\usepackage{mathtools}
\hypersetup{
    colorlinks=true,
    linkcolor=blue,
    filecolor=magenta,      
    urlcolor=blue,}

\newcommand{\LambdaQCD}{\Lambda_\text{QCD}}
\newcommand{\msbar}{\overline{\text{MS}}}

\journal{Nuclear Physics B}

\begin{document}

\begin{frontmatter}



\title{Precision Control in Lattice Calculation of $x$-dependent  \\Pion Distribution Amplitude}

\author[a,b]{Jack Holligan}
\ead{holligan@msu.edu}
\author[a]{Xiangdong Ji}
\ead{xji@umd.edu}
\author[b,c]{Huey-Wen Lin}
\ead{hueywen@msu.edu}
\author[a]{Yushan Su}
\ead{ysu12345@umd.edu}
\author[a]{Rui Zhang}
\ead{rayzhang@umd.edu, corresponding author}

\affiliation[a]{Department of Physics, University of Maryland, College Park, MD 20742, USA}
\affiliation[b]{Department of Physics and Astronomy, Michigan State University, East Lansing, MI 48824, USA}
\affiliation[c]{Department of Computational Mathematics, Science and Engineering, Michigan State University, East Lansing, MI 48824, USA }

\begin{abstract}
We present a new Bjorken $x$-dependence analysis of a previous lattice quantum chromodynamics data for the pion distribution amplitude from MILC configurations with three lattice spacing $a=0.06,0.09, 0.12$~fm. A leading renormalon resummation in renormalization as well as the perturbative matching kernel in the framework of large momentum expansion generates the power accuracy of the matching to the light-cone amplitude. Meanwhile, a small momentum log resummation is implemented for both the quark momentum $xP_z$ and the antiquark momentum $(1-x)P_z$ inside a meson of boost momentum $P_z$ up to 1.72 GeV along the $z$ direction, allowing us to have more accurate determination of the $x$-dependence in the middle range. Finally, we use the complementarity between the short-distance factorization and the large momentum expansion to constrain the endpoint regions $x\sim 0, 1$, thus obtaining the full-range $x$-dependence of the amplitude.

\end{abstract}




\end{frontmatter}


\section{Introduction}

Distribution amplitudes (DAs) are important observables for both theoretical and phenomenological reasons within the realm of quantum chromodynamics (QCD). The DA of a meson describes the probability amplitude of identifying the meson in a quark-antiquark Fock state on the lightcone, carrying longitudinal momentum fractions $x$ and $1-x$, respectively. It is also known as the leading Fock wave function of the meson. They are important as inputs to many exclusive processes with large momentum transfer, such as the B-meson decay, that can be factorized into the nonperturbative DA and the hard-scattering kernel~\cite{Beneke:1999br,Beneke:2001ev}. Although the DAs are important quantities in QCD, their properties, such as the moments, the shape and the endpoint power-law behavior are still undetermined from experiments~\cite{CLEO:1997fho,CELLO:1990klc,BaBar:2009rrj,Belle-II:2018jsg}. A direct nonperturbative calculation of the DAs from lattice QCD is thus of great interest.

The nonperturbative physics of partons is defined on the lightcone, i.e., in the effective limit of infinite momentum. Direct calculations on the lightcone are inaccessible on the lattice due to the dependence on real time. Early calculations determined DAs by calculating their lowest moments from local twist-2 operators~\cite{Kronfeld:1984zv,DelDebbio:2002mq,Braun:2006dg,Arthur:2010xf,Bali:2017ude,RQCD:2019osh} or from nonlocal current-current and quark bilinear correlators~\cite{Braun:2007wv,Braun:2015axa,Bali:2018spj,Detmold:2021qln,Gao:2022vyh}. The local-operator calculations provide precise measurements up to the second moment of the DA~\cite{RQCD:2019osh}, but the increasing noise and the nontrivial mixing in the lattice renormalization make it very difficult to access higher moments. The nonlocal-operator calculations analyze data in a certain current-current displacement or Wilson-line length $z$ range, where the short-distance factorization is valid; this either allows us to obtain the lowest few moments, or needs a model assumption to fit the $x$-dependence~\cite{Bali:2018spj,Gao:2022vyh}. A direct $x$-dependence calculation has not been possible in the two traditional methods.

The method of large-momentum effective theory (LaMET)~\cite{Ji:2013dva,Ji:2014gla,Ji:2020ect} offers a different approach, which starts from the Euclidean matrix element of equal-time, spatially separated fermion fields. After renormalization, we can physically extrapolate these matrix elements to large distances and Fourier transform them to momentum space. We use field-theoretical large momentum expansion to match the data at finite hadron-momentum to the light-cone distribution. This allows us to compute the precise shape of DA in the middle range of momentum fractions, $x$, without uncontrolled model fits. The first lattice calculation of the pion DA in LaMET was presented by the LP3 Collaboration in 2017~\cite{Zhang:2017bzy}, where they used a boosted pion at $P_z\approx 1.3$~GeV and a mass counterterm, $\delta m$, extracted from the Wilson-loop static potential to renormalize the matrix elements. A similar work on the kaon DA followed this work~\cite{Zhang:2017zfe}, studying its skewness and SU(3) symmetry, with a higher meson momentum $P_z\approx 1.7$~GeV and more precisely determined $\delta m$ by fitting to Wilson loops on multiple lattice spacings. The first continuum extrapolation $a\to 0$ was presented by the MSU group with nonperturbative renormalization~\cite{Zhang:2020gaj} on three lattice spacings down to $a\approx 0.06$~fm. In the latest calculation by LPC~\cite{Hua:2020gnw,LatticeParton:2022zqc}, the lattice artifacts have been controlled well by boosting to momentum as large as $P_z\approx 2.15$~GeV, extrapolating to the continuum, and calculating at the physical pion mass $m_\pi\approx 130$~MeV.

Now, the $x$-dependence calculation of the DA has entered an era of high precision, where the systematic controls on the theory side become more important. One complication in the calculation of quasi-DA matrix elements comes from the linearly divergent Wilson line self-energy~\cite{Ji:2017oey,Ishikawa:2017faj,Chen:2016fxx}. To remove the linear divergence without introducing extra nonperturbative effects, an improved renormalization in the hybrid scheme~\cite{Ji:2020brr} with self-renormalization~\cite{LatticePartonCollaborationLPC:2021xdx} at short distances has been applied to ensure the validity of the perturbative matching. However, such a renormalization is still not fully satisfactory, and there are more systematics to be controlled including the power accuracy, the large logarithms in the perturbative matching kernel and the endpoint region where the LaMET expansion breaks down.

The power accuracy is not automatically guaranteed in LaMET calculations, because of the ambiguity in renormalizing the linear divergence, and the factorially divergent coefficients in the perturbative matching. This ambiguity results in an extra $\mathcal{O}(\LambdaQCD/xP_z)$ leading power correction to the matching procedure. These have previously been handled by absorbing their behavior into a single fit parameter (often denoted $m_0$)~\cite{Gao:2021dbh,LatticeParton:2022zqc} when the lattice matrix elements are renormalized, while still using a fixed-order matching kernel. However, this method was examined in Ref.~\cite{lrr} and found to be inaccurate at twist-three level. The same paper describes a more robust method, known as leading-renormalon resummation (LRR), defining a renormalization scheme of the linear divergence in the perturbative results by resumming the contribution from the leading renormalon. Then the non-perturbative parameter $m_0$ can be extracted reliably to match the renormalized lattice data to $\overline{\rm MS}$ perturbative calculations to linear-$z$ accuracy, such that any linear-$z$ correction is eliminated. A corresponding LRR correction to the matching kernel eliminates the ambiguity from renormalization and thus improves the accuracy to $\mathcal{O}(\LambdaQCD/xP_z)$. We demonstrate in this work that the renormalization with LRR significantly improves the behavior of the quasi-DA short-distance correlations, making the results more consistent with the theoretical prediction from the operator product expansion (OPE).

The large momentum expansion of lattice observables includes both the nonperturbative light-cone distributions and the perturbative matching. The perturbative matching always involves logarithms of the renormalization scale $\mu$ and the physical scale of the system. When the logarithm becomes large, the higher-order effects are no longer negligible, and these higher-order logs need to be rearranged to ensure convergence of the perturbation series. This can usually be done by setting the renormalization scale $\mu$ equal to the physical one to eliminate the large logarithms, then solving the renormalization group (RG) equations to recover the $\mu$ dependence. This is known as renormalization-group resummation (RGR). When the physical scale becomes too small (around $\Lambda_\text{QCD}$), we reach the Landau pole, which suggests that perturbation theory breaks down or contains very large uncertainties. In the case of the parton distribution function (PDF), the only physical momentum is the parton momentum $xP_z$, so the logarithms become large at small $x$. Its effects and importance have been discussed in a recent work~\cite{Su:2022fiu}. The case of DAs is slightly different, as two different physical scales emerge. One scale is the quark momentum $xP_z$, and the other is the antiquark momentum $(1-x)P_z$. The existence of two different but correlated scales makes it more complicated to apply RGR because no choice of $\mu$ can eliminate the large logarithms simultaneously. In coordinate space, there is only one physical scale, the inverse of the correlation length $z^{-1}$. Thus, in principle it is easier to implement RGR in coordinate space for DAs before large momentum expansion. However, the coordinate-space matching is based on the short-distance twist expansion, which no longer works after reaching the Landau pole at $z\sim \Lambda_\text{QCD}^{-1}$. This prevents us from extracting the $x$ dependence of the DAs with coordinate-space matching. To obtain the lightcone DA with RGR, we propose an approach to resum the two logarithms simultaneously, which is valid in the mid-$x$ region, where LaMET gives accurate predictions.

In principle, the endpoint regions are not calculable from LaMET, because its applicable range is just the mid-$x$ region $x\in[x_\text{min},x_\text{max}]$ where $x_\text{min}\sim \LambdaQCD/P_z$ and $x_\text{max}=1-x_\text{min}$. However, the short-distance OPE provides us with global information about the lightcone distribution, allowing us to determine a few lowest moments of the lightcone DA, but not the local $x$-dependence. The two methods complement each other~\cite{Ji:2022ezo}, enabling us to combine the local information from the LaMET calculation and the global information from the short-distance OPE. We model the $x$-dependence outside the region $x\in [x_\text{min}, x_\text{max}]$ and fit to the short-distance correlations, obtaining a model-independent mid-$x$ distribution and a model-dependent endpoint distribution. The endpoint distribution is constrained by the physical requirement $\phi(x)\to 0$ when $x\to 0$ or 1 as well as the requirement of continuity with the LaMET calculation, which limit the model dependence. This combined approach provides a full $x$-dependence calculation.

With the above three ideas (elimination of the linear correction, resummation of large logarithms and constraining the endpoint regions), we improve the analysis of the lattice quasi-DA data to extract the full $x$-dependence of the lightcone DA with improved accuracy. The rest of the article is arranged as follows. In Sec.~\ref{sec:renormalon}, we describe the DA calculation using LaMET, discuss the ambiguities in the renormalization and perturbative matching, and present how we achieve power accuracy in the LaMET matching. In Sec.~\ref{sec:rgr_th}, we discuss the origin of two different physical scales in the LaMET matching and show how to resum them. In Sec.~\ref{sec:ope}, we discuss how to use the short-distance OPE to constrain the endpoint regions to extend our calculation to the full range of $x$. In Sec.~\ref{sec:numerical}, we apply LRR renormalization and LRR matching with RGR to extract the lightcone DA, then use complementarity to obtain the full $x$ dependence. Finally, we conclude in Sec.~\ref{sec:conclusion}. 


\section{Renormalization and Power Accuracy}\label{sec:renormalon}

The correlator that defines the pion DA on the lightcone is
\begin{align}\label{eq.LightconeDA}
\phi_{\pi}(x,\mu)= \int\frac{d \eta^-}{2\pi}e^{ixP^+\eta^-}\mel{0}{\overline{\psi}(0)\gamma_5\gamma_z W(0,\eta^-)\psi(\eta^-)}{\pi(P)},
\end{align}
where $W(0,\eta^-)=\hat{\mathcal{P}}\exp\left[-ig\int^{\eta^-}_0\!\!ds\,n_{\mu}A^{\mu}(ns)\right]$ is the Wilson line between the two points $0$ and $\eta^-$, and $\hat{\mathcal{P}}$ is the path-ordering operator. Lightcone coordinates are defined for a general Lorentz vector $V^{\mu}$, as $V^{\pm}=\frac{1}{\sqrt{2}}(V^0\pm V^3)$, since we may assume without loss of generality that the meson is traveling in the $z (\equiv\eta^3)$ direction. The term $g$ is the coupling, $A^{\mu}$ denotes the gauge field and $\ket{\pi(P)}$ is a pion state with 4-momentum $P^{+}$. The variable $x$ is the fraction of the meson momentum carried by the constituent parton.

The operator in Eq.~\eqref{eq.LightconeDA} has dependence on real time and is, thus, inaccessible directly on the lattice. The method of LaMET begins with the following ``quasi'' correlation:
\begin{align}
\label{eq.qLightconeDA}
    \tilde{\phi}_{\pi}(x,P_z)&=\int\frac{P_zdz}{2\pi}e^{ixP_zz}\tilde{h}^\text{R}_{\pi}(z,P_z)\\
    &=\int\frac{P_zdz}{2\pi}e^{ixP_zz}\mel{0}{\overline{\psi}(0,\vec{0})\gamma_5\gamma_z W(0,z)\psi(0,\vec{z})}{\pi(P_z)}_R,\nonumber
\end{align}
where $\tilde{h}^\text{R}_{\pi}$ is the renormalized coordinate-space matrix element defined in the second line. The lightcone DA, $\phi_{\pi}(x,\mu)$, is related to the quasi-DA (qDA) in the large momentum $P_z$ limit via
\begin{align}\label{eq.qDAandDA}
\tilde{\phi}_{\pi}(x,P_z) = \int^{1}_{0}dy\, \phi_{\pi}(y,\mu) \mathcal{C}(x,y,\mu,P_z) 
    +\mathcal{O}\left(\frac{\LambdaQCD^2}{x^2P^2_z},\frac{\LambdaQCD^2}{(1-x)^2P^2_z}\right),
\end{align}
where the $\mathcal{C}(x,y,\mu,P_z)$ is the perturbative matching kernel in momentum space, and the residual quadratic in $\LambdaQCD/P_z$ comes from higher-twist effects and is only leading if the linear divergence in the bare operator in Eq. (2) were not present as we explain below. 

The bare matrix elements we compute on the lattice are the $\tilde{h}^B_{\pi}(z,P_z,a)$ terms corresponding to Eq.~\eqref{eq.qLightconeDA} before renormalization, so our data are initially in position space and contain UV divergences. The spatial Wilson line $W(0,z)$ has a linearly-divergent self energy of size $\frac{1}{a}$, so besides the usual logarithmic divergence, the linear divergence must also be removed through a multiplicative renormalization~\cite{Stewart:2017tvs} before extrapolating to the continuum.
\begin{equation}\label{eq.Ratio}
    \tilde{h}^R(z,P_z,a) = \tilde{h}^B(z,P_z,a)/Z_R(z,a),
\end{equation}
where $Z_R(z,a) \sim e^{-\delta m(a) z}$ is the renormalization constant with the linearly divergent mass counterterm $\delta m(a)\sim\frac{1}{a}$.

When renormalizing the linear divergence, one could in principle also choose to subtract a finite constant term along with it. The choice of this finite piece defines the renormalization scheme. Also, when expanding $\delta m(a)$ as a perturbation series in the strong coupling, $\alpha_s$,
\begin{equation}
\delta m = \frac{1}{a} \sum_n r_n \alpha_s^{n+1}(a^{-1}),
\end{equation}
the coefficient $r_n\sim n!$ grows factorially at higher orders due to an infrared renormalon effect~\cite{Beneke:1998ui,Bali:2013pla}. Thus, the series is divergent for any $\alpha_s$, and the sum is ill-defined. To fix this degree of freedom, we need to introduce an additional renormalization scheme for the linear divergence to define $\delta m (a,\tau)$ unambiguously, with a new $\tau$-dependence. This result varies by $\mathcal{O}(\LambdaQCD)$ in different $\tau$-schemes, so an ambiguity of $\mathcal{O}(z\LambdaQCD)$ arises in the renormalization factor. The same intrinsic ambiguity appears when we try to extract $\delta m$ from fitting lattice data, where $\delta m$ is always mixed with another non-perturbative quantity, such that we have a freedom to choose among different fitting results. Similarly, a calculation of the perturbative matching kernel $\mathcal{C}(x,y,\mu,P_z)$ also suggests a factorial growth with the same pattern~\cite{Braun:2018brg}. The lightcone distribution $\phi(x,\mu)$ is obtained by convoluting the inverse matching kernel $\mathcal{C}^{-1}(x,y,\mu,P_z)$ with the renormalized quasi-DA, $\tilde{\phi}_{\pi}(x,P_z)$, both containing the ambiguities. The combination will, in general, result in a linear correction $\mathcal{O}\left(\frac{\LambdaQCD}{xP_z}\right)$ to the matching~\cite{Ji:2020brr},
\begin{align}\label{eq.matching_with_correction}
\tilde{\phi}_{\pi}(x,P_z) = \int^{1}_{0}dy\, \phi(y,\mu) \mathcal{C}(x,y,\mu,P_z) +\mathcal{O}\left(\frac{\LambdaQCD}{xP_z}\right) +\mathcal{O}\left(\frac{\LambdaQCD^2}{x^2P^2_z}\right),
\end{align}
where we have ignored the $(1-x)P_z$ scale for simplicity, which can be recovered by a substitution $x\leftrightarrow 1-x$, due to the symmetry of the matching. When the hadron momentum is large enough, this correction is not important. But the hadron states in lattice calculations are usually moving with $P_z\sim$~GeV, where the linear correction can be large, especially near the endpoints, and more important than the quadratic higher-twist effects.

In principle, these ambiguities from the renormalization and the perturbative matching can cancel because the twist-2 lightcone DA $\phi(x,\mu)$ is free of the linear divergence and the infrared renormalon. Thus, the power accuracy up to $\mathcal{O}\left(\frac{\LambdaQCD}{xP_z}\right)$ is the best we can achieve without knowing higher-twist information. It is shown that they indeed cancel only when the renormalization of linear divergence and the regularization of the matching coefficients are defined in the same $\tau$-scheme~\cite{lrr}. To achieve this accuracy, we need to carefully define the renormalization scheme for the linear divergence $\delta m$ and regularize the perturbative matching consistently to eliminate the linear correction.

A recent work~\cite{LatticeParton:2022zqc} uses a fixed-order approach to handle this ambiguity by introducing an additional twist-three mass parameter, denoted by $m_0$, in the renormalization process to ensure that the short-distance behavior of the renormalized matrix element is in agreement with perturbation theory. The approach is still not good enough for several reasons: 1) A bridge is missing to connect the lattice calculation and the perturbative calculation, usually known as the scheme-conversion factor in the renormalization; 2) Resumming the logarithms $\ln(z^2\mu^2)$ at short distances clearly suggests that $m_0$ is not a constant but has a large dependence on $z$, mainly due to the fixed-order truncation not being a proper scheme to regularize the divergent series~\cite{lrr}; 3) A fixed-order matching is used, which cannot eliminate the linear ambiguity.

We propose a new approach, aimed at eliminating such a correction to achieve the power accuracy, as demonstrated in Ref.~\cite{lrr}. It includes four steps:
\begin{itemize}
    \item Modify the perturbative matching coefficients through a leading renormalon resummation (LRR) with a principal value (PV) prescription, defined as the $\tau$-scheme. 
    \item Determining the non-perturbative twist-3 parameter $m_0(\tau)$ through the matching condition that the renormalized $P_z=0$ lattice data agree with the LRR-improved Wilson coefficients in the $\tau$-scheme up to twist-3 accuracy in the OPE at short distances;
    \item Renormalize the $P_z>0$ lattice data with the $m_0(\tau)$ extracted from the previous step in the $\tau$-scheme;
    \item Extract the DA with the LRR-improved perturbative matching kernel.
\end{itemize}

The $m_0(\tau)$ parameter is fixed by $P_z=0$ pion quasi-PDF data and used for the renormalization of the quasi-DA at nonzero momentum. The justification for this choice is that the linear correction from the ambiguity in the linear divergence is independent of the momentum of the external state and the Dirac structure. $\delta m$ is universal for the Wilson-line self energy, which is the same in these observables ($P_z=0$ and $P_z> 0$) obtained from the same gauge action; the leading renormalon contribution that we resum in the $\msbar$ perturbative calculation also originates from the Wilson line self energy, thus is the same for these observables, up to an overall phase factor depending on the external states' momentum. Thus, once the cancellation of the ambiguities is achieved for one observable, it is also guaranteed for other observables of a similar structure, i.e., with the same Wilson line in the quark bilinear operator but with a different Dirac structure or different external states.

The LRR improves the renormalization method in two aspects, as we will show in Sec.~\ref{sec:numerical}. Firstly, the $e^{m_0(\tau)z}$ factor is extracted with LRR, so it is different from that extracted in fixed-order perturbation theory. With LRR improvement, its extraction is almost independent of small-$z$ values, and is determined with a significantly reduced uncertainty from scale variation~\cite{lrr}. Secondly, in the hybrid scheme, the renormalized matrix elements are divided by $P_z=0$ perturbative results (i.e., the Wilson coefficient $C_{00}$) at short distances, whose $z$-dependence is improved after LRR in the sense that they are more consistent with the OPE at short distances, with the lowest few moments as inputs. This moment is supposed to be consistent with the one extracted from a renormalization-independent ratio between two different momenta, which will be discussed in more detail in Sec.~\ref{sec:ope}. Thus, a comparison between the moments extracted from the renormalization-dependent matrix element and from the renormalization-independent ratio will test whether the renormalization is properly done. 

The idea of LRR is to resum the leading factorially divergent high-order terms to all orders in the perturbation series. Then the remaining part of the series, if without other renormalons, is convergent, and any leading power correction could be fixed by a regularization of the resummed divergent part. Although it is impossible to analytically calculate the perturbation for specific processes to all orders, we can calculate a specific type of bubble-chain diagrams~\cite{tHooft:1977} in the large $\beta_0$ limit. Beyond the large-$\beta_0$ limit, the asymptotic form of the leading renormalon pole is known~\cite{Beneke:1998ui,Pineda:2001zq}, whose overall strength has been estimated from perturbation series of the heavy quark pole mass~\cite{Pineda:2001zq,Pineda:2002se} and lattice calculations of the static potential~\cite{Bali:2013pla}. Thus we can also choose to resum these known asymptotic forms. These two approaches both resum the leading pole corresponding to the linear divergence, but have different ``background'' effects that are higher powers of $\Lambda_{\rm QCD}$ and higher order of $\alpha_s$. Thus they are supposed to make slightly different predictions in the mid-$x$ region, as we will discuss in Sec.~\ref{sec:numerical}.

\subsection{LRR in the large $\beta_0$ limit}
For quasi-PDF operators, a calculation for bubble-chain diagrams has been done in Ref.~\cite{Braun:2018brg}. Note that only the Wilson-line self-energy diagram (also called the ``tadpole'' diagram in Ref.~\cite{Izubuchi:2018srq}) is relevant to the leading renormalon, so we can ignore the other diagrams which only account for higher renormalon poles.

By resumming the tadpole diagrams, the LRR in the large-$\beta_0$ limit modifies the $P_z=0$ matrix element for the DA, i.e., the Wilson coefficient $C_{00}(z,\mu)$, in the following way:
\begin{align}
\label{eq:me_pv}
    C_{00}^\text{LRR}(z,\mu,\tau)= C_{00}^\text{tp}(z,\mu)|_\text{PV}+\sum_i\left( C_{00}^{(i)}(z,\mu)-C_{00}^{\text{tp},{(i)}}(z,\mu)\right),
\end{align}
where $C_{00}^\text{tp}(z,\mu)|_\text{PV}$ is the resummed diagrams with the principal value prescription for the poles defined as scheme $\tau$,
\begin{align}
\label{eq:borel_integral}
    C_{00}^\text{tp}(z,\mu)|_\text{PV}=\int_\text{0, PV}^{\infty}du  e^{-4\pi u/\alpha(\mu)\beta_0} \frac{2C_F}{\beta_0} \times \left(\frac{\Gamma(1-u)e^{\frac{5}{3}u}(z^2\mu^2/4)^u}{(1-2u)\Gamma(1+u)}-1\right)/u,
\end{align}
and $C_{00}^{\text{tp},{(i)}}(z,\mu)$ is the $i$-th order expansion of $C_{00}^\text{tp}(z,\mu)|_\text{PV}$ in $\alpha_s$. At NLO, we have
\begin{align}
    C_{00}^{\text{tp},{(1)}}(z,\mu)=\frac{\alpha_sC_F}{2\pi}\left(\ln\left(z^2\mu^2e^{2\gamma_E}/4\right)+\frac{11}{3}\right),
\end{align}
and the corresponding Wilson coefficient~\cite{Izubuchi:2018srq}
\begin{align}
    C_{00}^{(1)}(z,\mu)=\frac{\alpha_sC_F}{2\pi}\left(\frac{3}{2}\ln\left(z^2\mu^2e^{2\gamma_E}/4\right)+\frac{7}{2}\right).
\end{align}
The $P_z>0$ matrix elements $\tilde{H}(z,P_z,\mu)$ are corrected by LRR in a similar way, where the momentum dependence only enters through a phase factor,
\begin{align}
\label{eq:me_pv}
    \tilde{H}^\text{LRR}(z,yP_z,\mu,\tau)=&\tilde{H}(z,yP_z,\mu)+e^{-iyzP_z}C_{00}^\text{tp}(z,\mu)|_\text{PV}-\sum_ie^{-iyzP_z}C_{00}^{\text{tp},{(i)}}(z,\mu),
\end{align}
which can be Fourier transformed to obtain the correction to the NLO matching kernel in $\msbar$,
\begin{align}
    \Delta\mathcal{C}^{\msbar}(x,y,P_z,\mu,\tau)
    &=\int \frac{P_zdz}{2\pi} e^{i(x-y)zP_z} \left(C_{00}^\text{tp}(z,\mu)|_\text{PV}-C_{00}^{\text{tp},{(1)}}(z,\mu)\right)\nonumber\\
    &=\int \frac{P_zdz}{2\pi} e^{i(x-y)zP_z} C_{00}^\text{tp}(z,\mu)|_\text{PV}-\frac{\alpha_sC_F}{2\pi}\frac{1}{|x-y|}\nonumber,
\end{align}
where the first part is not a traditional convergent function, but a distribution operating on the DA function through a convolution, whose effect is convergent mathematically. In practice, it is enough to perform a truncated numerical evaluation to some large $z_\text{max}$, e.g., $z_\text{max}=10$~fm.

In the ratio scheme~\cite{Radyushkin:2017cyf,Orginos:2017kos,Radyushkin:2017lvu}, the ratio between two momentums is free of linear divergence, thus no LRR modification is needed, and the matching kernel $\mathcal{C}^\text{ratio}$ is unchanged.
In the hybrid scheme, the correction is an integration from $z_s$ to $z_{\rm max}$ during the Fourier transformation,
\begin{align}
    &\Delta\mathcal{C}^\text{hybrid}(x,y,P_z,\mu,\tau)\nonumber\\
    =&2\int_{z_s}^\infty \frac{P_zdz}{2\pi} \cos\left((x-y)zP_z\right) \left(C_{00}^\text{tp}(z,\mu)|_\text{PV}
   -C_{00}^{\text{tp},{(1)}}(z,\mu)-C_{00}^\text{tp}(z_s,\mu)|_\text{PV}+C_{00}^{\text{tp},{(1)}}(z_s,\mu)\right)\nonumber\\
    =&2\int_{z_s}^{z_\text{max}} \frac{P_zdz}{2\pi} \cos\left((x-y)zP_z\right) (C_{00}^\text{tp}(z,\mu)|_\text{PV}-C_{00}^\text{tp}(z_s,\mu)|_\text{PV})+\frac{\alpha_sC_F}{2\pi}\left(\frac{1}{|x-y|}-2\frac{\text{Si}\left((x-y)z_sP_z\right)}{\pi(x-y)}\right),
\end{align}
where the first term can be calculated numerically to $z_\text{max}$ in practice.

It is also straightforward to derive the LRR correction to the DA Wilson coefficients by expanding Eq.~\eqref{eq:me_pv} in $zP_z$:
\begin{align}
    \Delta C_{kl}^\text{LRR}=\delta_{kl} \left(C_{00}^\text{tp}(z,\mu)|_\text{PV}-\sum_iC_{00}^{\text{tp},{(i)}}(z,\mu)\right),
\end{align}
which can be applied to the OPE of short distance correlations.

\subsection{LRR of the asymptotic series}
Besides resumming the leading renormalon pole, the large-$\beta_0$ approximation introduces extra effects in subleading renormalon poles. Alternatively, as discussed in Ref.~\cite{lrr}, we can resum the asymptotic form of the leading renormalon contribution, which only includes the leading renormalon pole. In this approach, we utilize the fact that the leading renormalon contribution originates from the heavy quark pole mass $m = \mu \sum_n r_n\alpha_s^{n+1}$, with a known asymptotic form in large perturbation order $n$~\cite{Beneke:1998ui,Pineda:2002se,Bali:2013pla}, 
\begin{align}\label{eq:asymptotic_form_2}
	r_n = N_m \left(\frac{\beta_0}{2\pi}\right)^n
	\frac{\Gamma(n+1+b_0)}{\Gamma(1+b_0)}
	\left[1+\frac{c_1b_0}{b_0+n} +...\right],
\end{align} 
where $b_0=\beta_1/2\beta_0^2$ and $c_1=(\beta_1^2-\beta_0\beta_2)/(4b_0\beta^4_0)$ are from higher orders in the QCD beta function. Using an analytical method in Ref.~\cite{Pineda:2001zq}, the overall strength can be determined as $N_m(n_f=3) =0.575$, $N_m(n_f=4) =0.552$. Thus the contribution to the DA Wilson coefficients has the following form at large $n$:
\begin{align}
    C^{(n+1)}_{kl}(z,\mu)\xrightarrow{n\to\infty}\delta_{kl} z\mu r_n\alpha_s^{n+1}(\mu).
\end{align}
Similar to the LRR in the large-$\beta_0$ limit, we can resum the asymptotic form with the PV prescription,
\begin{align}
	\label{eq:borel_integral}
	C^{\rm asympt}_{kl}(z,\mu)_{\rm PV}= \delta_{kl}N_m z\mu\frac{4\pi}{\beta_0}   \int_{\rm 0, PV}^{\infty}du   e^{-\frac{4\pi u}{\alpha_s(\mu)\beta_0}}  \frac{1}{(1-2u)^{1+b_0}}\big(1+c_1(1-2u)+...\big).
\end{align}
It's easy to verify that the ambiguity of this integral is linear in $z\Lambda_{\rm QCD}$ and independent of $\mu$. Note that the Fourier transformation of this correction can be calculated analytically, but the explicit linear-$z$ dependence will be transformed into a singular distribution of $x-y$, including derivatives of the $\delta(x-y)$ function. It is numerically very unstable if this function is applied to discrete data. So a regularization is applied, by multiplying the linear-$z$ term with a small exponential decaying factor $\exp(-\epsilon_m z)$, which will result in extra higher-twist corrections $\mathcal{O}\left(\frac{\epsilon_m\LambdaQCD}{4x^2P^2_z}\right)$ that is insignificant in mid-$x$ region. With such a regularization, the correction to the hybrid-scheme matching will be
\begin{align}
	\Delta& \mathcal{C}^{\rm hybrid}(\Delta_{xy},\mu,P_z,\tau)=(C^{\rm asympt}_{kl}(z,\mu)_{\rm PV}/z-r_0\mu\alpha_s(\mu))\left\{\frac{e^{-\epsilon_m z_s}P_z(1+\epsilon_m z_s+\epsilon_m^2z_s^2)}{\epsilon_m^2\pi}+\frac{1}{\pi}\left(\frac{e^{-\epsilon_m z_s}z_s(\sin[\Delta_{xy} z_sP_z])}{\Delta_{xy}}\right.\right.\\
	+&\left.\left.\frac{e^{-\epsilon_m z_s}P_z}{(\epsilon_m^2+P_z^2\Delta_{xy}^2)^2}\left((\epsilon_m^2-\Delta_{xy}^2P_z^2+\epsilon_m^3z_s+\epsilon_mP_z^2\Delta_{xy}^2z_s)\cos[\Delta_{xy} z_sP_z]-\Delta_{xy} P_z(2\epsilon_m+\Delta_{xy}^2P_z^2z_s+\epsilon_m^2z_s)\sin[\Delta_{xy} z_sP_z]\right)\right)\right\}_+.\nonumber
\end{align}
where $\Delta_{xy}=|x-y|$. The overall factor $C^{\rm asympt}_{kl}(z,\mu)_{\rm PV}/z-r_0\mu\alpha_s(\mu)$ only depends on $\mu$ and can be integrated numerically. The total correction is written as a plus function to guarantee the current conservation because one term proportional to $\delta(x-y)$ has been omitted. Testing with some different $\epsilon_m\sim20-100$ MeV values, and with $\Delta x=0.01$ as the step size of our numerical methods in the momentum space matching, we find the results are consistent and stable. Working with smaller $\epsilon_m$ requires a finer discretization of the data as a function of $x$ or $y$.


\section{Small-momentum large logarithm resummation}\label{sec:rgr_th}


\subsection{Resummation in coordinate space}\label{sec:resummation}

To study the resummation of large logarithms, we start from a simpler case, the coordinate-space matching of the quasi-DA. It is more straightforward because only one physical scale, $z=\lambda/P_z$, is involved in the matching. The renormalized quasi-DA matrix element, $\tilde{h}^{\rm R}(\lambda,P_z)$, can be matched to lightcone DA, $h(\lambda,\mu)$, through
\begin{equation}
\label{eq:sdf}
\tilde{h}^{\rm R}(\lambda,P_z) = \int_0^1 d\nu\, h(\nu\lambda,\mu) \mathcal{Z}(\nu,z^2,\mu^2,\lambda) + \mathcal{O}(z^2\Lambda_\text{QCD}^2),
\end{equation}
where $\mathcal{Z}(\nu,z^2,\mu^2,\lambda)$ is the perturbative matching kernel in coordinate space. In the ratio scheme~\cite{Radyushkin:2017cyf,Orginos:2017kos,Radyushkin:2017lvu},
\begin{align}
\label{eq:corr_matching}
    \mathcal{Z}&(\nu,z^2,\mu^2,\lambda)=\delta(1-\nu)
    +\frac{\alpha_sC_F}{2\pi}\left\{\left(\frac{\nu}{1-\nu}\right)_+\left(-(1+L)(1+e^{-i\lambda(1-\nu)})\right)\right.\nonumber\\
    &-2\left(\frac{\ln(1-\nu)}{1-\nu}\right)_+(1+e^{-i\lambda(1-\nu)})
    \left.+\left(\frac{1-e^{-i\lambda(1-\nu)}}{i\lambda}-\frac{1}{2}\delta(1-\nu)\right)(3-L)\right\},
\end{align}
where $L=\ln\left(z^2\mu^2e^{2\gamma_E}/4\right)$ is the only scale-dependent logarithm appearing in the kernel and $C_F$ is the quadratic Casimir for the fundamental representation of SU(3).
At either short distances, $z\to0$, or long distances, $z\gg \LambdaQCD^{-1}$, the logarithm becomes large. We can eliminate the logarithm by setting $\mu_0=2e^{-\gamma_E}z^{-1}$ on the right-hand side of the equation. Then an RG evolution to the default scale, e.g. $\mu=2$~GeV, will resum the large logarithms at that scale,
\begin{equation}
\tilde{h}^{\rm R}(\lambda,P_z)
    =\int_0^1d\nu \exp[\int_{\mu}^{\mu_0}\hat{\mathcal{V}}(\rho/\nu,\lambda,\mu')d\ln\mu'^2]h(\rho\lambda,\mu)\mathcal{Z}(\nu,z^2,\mu_0^2,\lambda)+\mathcal{O}(z^2\Lambda_\text{QCD}^2),
\end{equation}
where $\hat{\mathcal{V}}$ is the coordinate space representation of the Efremov-Radyushkin-Brodsky-Lepage (ERBL) evolution kernel~\cite{Efremov:1978rn,Efremov:1979qk,Lepage:1979zb,Lepage:1980fj},
\begin{equation}
\hat{\mathcal{V}}(\nu,\lambda,\mu) = \frac{\alpha_s(\mu)C_F}{2\pi}\left((1+e^{-i\lambda(1-\nu)})\left(\frac{\nu}{1-\nu}\right)_+ +\left(\frac{1-e^{-i\lambda(1-\nu)}}{i\lambda}-\frac{1}{2}\delta(1-\nu)\right)\right)\theta(1-\nu).
\end{equation}
Both the evolution and matching kernels can be made purely real by multiplying by a phase factor $e^{i\lambda(1-\nu)/2}$, and applying these to the phase-rotated matrix elements
\begin{align}
\label{eq:phase_rot}
\tilde{H}^R(\lambda,P_z)=e^{i\lambda/2}\tilde{h}^R(\lambda,P_z),
\end{align}
which is purely real for symmetric DAs that satisfy $\phi(x)=\phi(1-x)$. So the matching and the evolution preserve the symmetry of the DAs.

Such a resummation works fine at short distances, but at long distances the scale $\mu_0$ hits the Landau pole, indicating that the perturbation theory as well as the entire short-distance operator expansion break down. Without knowing the correct information at large $z$, we are unable to extract the $x$ dependence of the DA.


\subsection{Origin of two different scales}

In large momentum expansion, we perform the resummation and matching in momentum space, and the higher-twist non-perturbative physics appear now at the endpoint regions $x\to 0$ and $x\to 1$, which we will choose to model using complementarity, as we will discuss in the next section. In momentum space, the quasi-DA is matched to the lightcone DA through
\begin{align}
\label{eq:matching_mom}
\tilde{\phi}(x,P_z) =
\int_0^1 dy\, \mathcal{C}(x,y,\mu,P_z)\phi(y,\mu) + \mathcal{O}\left(\frac{\Lambda^2_\text{QCD}}{y(1-y)P^2_z}\right).
\end{align}
The momentum-space matching kernel $\mathcal{C}(x,y,\mu,P_z)$ can be obtained from a double Fourier transformation of the coordinate-space matching in Eq.~\eqref{eq:corr_matching},
\begin{align}
\mathcal{C}(x,y,\mu,P_z) = 
\int^{\infty}_{-\infty}\frac{d\lambda}{2\pi}e^{ix\lambda} \int_0^1 d\nu\, e^{-i\nu y\lambda} \mathcal{Z}\left(\nu,\frac{\lambda^2}{y^2P_z^2},\mu^2\right).
\end{align}
To trace how the physical scale and the logarithm transform, we can check the double Fourier transform of the logs, $L$, in $\mathcal{Z}(\nu,\frac{\lambda^2}{y^2P_z^2},\mu^2)$ of Eq.~\eqref{eq:corr_matching}. The terms involved include $f(\nu)L$ and $f(\nu)Le^{-i\lambda(1-\nu)}$. In dimensional regularization $d=4-2\epsilon$, higher-order logs $L^n$ can be expressed as the $\mathcal{O}(1)$ term of $\frac{n!}{\epsilon^n}\left(\mu^2z^2e^{2\gamma_E}/4\right)^\epsilon$ in the $\epsilon$ expansion.
Integrating $\lambda$ first, we obtain
\begin{align}
    &\int^{\infty}_{-\infty} \frac{d\lambda}{2\pi} e^{ix\lambda}\int_0^1 d\nu\,e^{-i\nu y\lambda}\frac{n!f(\nu)}{\epsilon^n}\left(\frac{\lambda^2\mu^2e^{2\gamma_E}}{4P_z^2}\right)^\epsilon
    \nonumber\\
    &=\int_0^1\frac{d\nu f(\nu)n!}{|x-y\nu|^{1+2\epsilon}}\frac{\Gamma(1/2+\epsilon)}{\sqrt{\pi}\epsilon^n\Gamma(-\epsilon)}\left(\frac{\mu^2e^{2\gamma_E}}{P_z^2}\right)^\epsilon\nonumber\\
    &=\int_0^1\frac{d\nu f(\nu)n!}{y|x/y-\nu|^{1+2\epsilon}}\frac{\Gamma(1/2+\epsilon)}{\sqrt{\pi}\epsilon^n\Gamma(-\epsilon)}\left(\frac{\mu^2e^{2\gamma_E}}{y^2P_z^2}\right)^\epsilon,
\end{align}
which can be expanded in $\epsilon\to 0$~\cite{Izubuchi:2018srq}. Note that only when $0<x/y<1$ is it possible for $|x/y-\nu|$ to be zero in the integration region $\nu\in[0,1]$, and then the expansion of $|x/y-\nu|^{-1-2\epsilon}$ generates the leading divergent term $\frac{-1}{2\epsilon}\delta(x/y-\nu)$. So when $x$ is in the nonphysical region, or when $1>x>y>0$, the expansion does not contain any leading logarithms of $(\ln\mu)^n$. When $0<x<y<1$, the expansion yields additional log terms
\begin{align}
&\int_0^1 d\nu f(\nu) \left(\ln^n (\mu^2)-n\ln^{n-1}(\mu^2)\ln(4y^2P_z^2)\right)
\times\delta(x-y\nu)  + \mathcal{O}(\ln^{n-2}\mu)\nonumber\\
&=\int_0^1 d\nu f(\nu)\delta(x-y\nu) \ln^n\left(\frac{\mu^2}{4y^2P_z^2}\right) + \mathcal{O}(\ln^{n-2}\mu).\nonumber
\end{align}
The remaining integral preserves the structure of the log and only changes its coefficients. Thus, we get the physical scale $2yP_z$ for this term.
On the other hand, the other term $f(\nu)Le^{-i\lambda(1-\nu)}$ after a double Fourier transformation becomes
\begin{align}
     &\int \frac{d\lambda}{2\pi}  e^{ix\lambda}\int_0^1 d\nu\,e^{-i\nu y\lambda}e^{-i\lambda(1-\nu)}\frac{n!f(\nu)}{\epsilon^n}\left(\frac{\lambda^2\mu^2e^{2\gamma_E}}{4P_z^2}\right)^\epsilon
    \nonumber\\
    &=\int_0^1\frac{d\nu f(\nu)n!}{|\overline{x}-\overline{y}\nu|^{1+2\epsilon}}\frac{\Gamma(1/2+\epsilon)}{\sqrt{\pi}\epsilon^n\Gamma(-\epsilon)}\left(\frac{\mu^2e^{2\gamma_E}}{P_z^2}\right)^\epsilon\nonumber\\
    &=\int_0^1\frac{d\nu f(\nu)n!}{\overline{y}|\overline{x}/\overline{y}-\nu|^{1+2\epsilon}}\frac{\Gamma(1/2+\epsilon)}{\sqrt{\pi}\epsilon^n\Gamma(-\epsilon)}\left(\frac{\mu^2e^{2\gamma_E}}{\overline{y}^2P_z^2}\right)^\epsilon,
\end{align}
with $\overline{x}\equiv1-x$ and $\overline{y}\equiv1-y$, which does not contain any leading logarithm when $1-x$ is nonphysical, or when $0<x<y<1$. When $1>x>y>0$, the $\epsilon$ expansion yields
\begin{align}
    \int_0^1 d\nu f(\nu) \delta(\overline{x}-\overline{y}\nu) \ln^n\left(\frac{\mu^2}{4\overline{y}^2P_z^2}\right)
    +\mathcal{O}(\ln^{n-2}\mu).
\end{align}
So the two different physical scales correspond to different regions of $x$ and $y$. The scale $2yP_z$ for $x<y$ corresponds to the quark-splitting process; the scale $2(1-y)P_z$ for $x>y$ corresponds to the antiquark-splitting process.

Note that the two scales $2yP_z$ and $2(1-y)P_z$ we obtained at the current stage both depend on the convolution variable $y$, which is, in principle, not implementable because we cannot have different scales $\mu$ for different $y$ in $\phi(\mu,y)$. However, note that the matching kernel $\mathcal{C}(x,y,\mu,P_z)$ is almost localized, i.e., the region of $x\sim y$ is greatly enhanced compared to any other regions. As a result, the proper scale choice to resum the RG logarithm would be $2xP_z$ and $2(1-x)P_z$ instead, corresponding to the quark and antiquark momentum fractions in the quasi-DA. Moreover, a comparison between the quasi-PDF and DIS shown in Ref.~\cite{Su:2022fiu} suggests that $2xP_z$ is the proper scale in the quasi-PDF case, also supports the scales to be $2xP_z$ and $2(1-x)P_z$ in our quasi-DA. We can examine the sensitivity to this choice of scale by slightly varying the value from $2xP_z$ to $2cxP_z$ with $c\in[0.75,1.5]$, which roughly correspond to a $\pm30\%$ change in $\alpha_s$ near $1$~GeV.


\subsection{Resummation of two different logarithms}

The perturbative matching kernel in Eq.~\eqref{eq:matching_mom} $\mathcal{C}(x,y,\mu,P_z)=\delta(x-y)+\mathcal{C}^{(1)}(x,y,\mu,P_z)+\mathcal{O}(\alpha_s^2)$ has been calculated to 1-loop order~\cite{Liu:2018tox}, where $\mathcal{C}^{(1)}$ is the $\mathcal{O}(\alpha_s)$ term of the matching, and satisfies the quark-antiquark symmetry
\begin{equation}
    \mathcal{C}(x,y,\mu,P_z) = \mathcal{C}(1-x,1-y,\mu,P_z).
\end{equation}
Moreover, it contains logarithms of both kinds as discussed in the previous subsection, which becomes large at the end-point regions.  Indeed, in the two regions $x<y$ and $y<x$, the matching kernel has different $\mu$ dependencies, corresponding to the piecewise function of the ERBL evolution kernel:
\begin{equation}
    \frac{d\phi(x)}{d\ln\mu^2} = \int_0^1dy \frac{\alpha_S C_F}{2\pi}V^{(0)}(x,y)\phi(y) + \mathcal{O}(\alpha^2_S),
\end{equation}
where
\begin{equation}
    V^{(0)}(x,y) = \left(\frac{x}{y}\frac{1-x+y}{y-x}\theta(y-x)+\left[\begin{matrix}
    x\leftrightarrow\overline{x}\\y\leftrightarrow\overline{y}\end{matrix}
    \right]\right)_+,
\end{equation}
and the plus function is
\begin{align}
    f(x,y)_+=f(x,y)+\delta(x-y)\int_0^1 dz\,f(z,y).
\end{align}
The logarithms become large in the matching kernel $\mathcal{C}(x,y,\mu,P_z)$ for $x$ close to both endpoints $x\to 0$ or 1, so a resummation of large logs is necessary. 

The traditional method of resummation is to choose a scale $\mu_0$ in the scale-independent factorization such that the large logs in $\mathcal{C}(x,y,\mu_0,P_z)$ are eliminated, and apply the RG evolution of $\phi(y,\mu)$:
\begin{equation}
    \tilde{\phi}(x,P_z) = \int_0^1 dy\, \mathcal{C}(x,y,\mu_0,P_z) \mathcal{P}\exp\left[\int_\mu^{\mu_0} \hat{V}d\ln\mu'^2\right]\phi(y,\mu),
\end{equation}
where $\mathcal{P}$ is the path ordering of the evolution path, and $\hat{V}$ is the operator corresponding to the ERBL kernel acting on $\phi$. However, such an approach does not work in the quasi-DA's case, because there are two different scales in the piecewise matching kernel. The logarithm in $x<y$ is $\ln[4x^2P_z^2/\mu^2]$, corresponding to the quark-splitting; the logarithm in $x>y$ is $\ln[4(1-x)^2P_z^2/\mu^2]$, corresponding to the antiquark-splitting. No single choice of $\mu_0$ is able to eliminate the two logs at the same time. We have to develop a different strategy to resum the log in the DA.

We start from separating the matching formula in two different regions
\begin{align}
    \tilde{\phi}(x)&= \phi(x,\mu)+\int_{\mathrlap{x<y}}\,dy\, \mathcal{C}^{(1)}(x,y,\mu)\phi(y,\mu)
    +\int_{\mathrlap{x>y}}\,dy\,\mathcal{C}^{(1)}(x,y,\mu)\phi(y,\mu)+\mathcal{O}(\alpha^2_s)\nonumber\\
    &=\phi(x,\mu)+\mathcal{C}^{(1)}_L(\mu)\otimes\phi(\mu)+\mathcal{C}^{(1)}_R(\mu)\otimes\phi(\mu),
\end{align}
where we label the integral in two regions as $\mathcal{C}^{(1)}_L$ and $\mathcal{C}^{(1)}_R$ convoluted with $\phi$. One idea is to set the two terms to different scales, $\mathcal{C}^{(1)}_L(2xP_z)\otimes\phi(2xP_z)$ and $\mathcal{C}^{(1)}_R(2(1-x)P_z)\otimes\phi(2(1-x)P_z)$. To make this possible, we need to split the first term into two parts and combine with them separately. After the split, the two parts need to be scale-invariant individually up to order $\alpha_s$. So we need to find the weights, $w_L(x)$ and $w_R(x)=1-w_L(x)$, for the following split
\begin{align}
    \tilde{\phi}(x)&=w_L(x)\phi(x,\mu)+\mathcal{C}^{(1)}_L(\mu)\otimes\phi(\mu)\\
    &+w_R(x)\phi(x,\mu)+\mathcal{C}^{(1)}_R(\mu)\otimes\phi(\mu),
\end{align}
such that the two lines are individually scale-independent. If we force $w_L(x)$ to be scale independent, then simple algebra gives
\begin{align}
\label{eq:sol_mu}
    w_L(x)=\frac{V^{(0)}_L\otimes\phi(\mu)}{V^{(0)}\otimes\phi(\mu)},
\end{align}
where we define $V^{(0)}_L\equiv V^{(0)}|_{x<y}$ and $V^{(0)}_R\equiv V^{(0)}|_{x>y}$ in a similar way. The solution $w_L(x)$ is scale-dependent at order $\alpha_s$, contradicting our requirement. However, noticing that $\phi(\mu)$ and $\tilde{\phi}$ differ by $\mathcal{O}(\alpha_s)$ corrections, and the quasi-DA $\tilde{\phi}$ is scale independent, we may substitute the $\phi(\mu)$ in Eq.~\eqref{eq:sol_mu} with $\tilde{\phi}$, making $w_L(x)$ scale independent. The new solution
\begin{align}
\label{eq:sol_no_mu}
    w_L(x)=\frac{V^{(0)}_L\otimes\tilde{\phi}}{V^{(0)}\otimes\tilde{\phi}}
\end{align}
satisfies the scale dependence of the original equation at order $\mathcal{O}(\alpha_s)$,
\begin{align}
    \frac{d w_L(x)\phi(x)}{d\ln\mu^2}&=\frac{\alpha_s C_F}{2\pi}\frac{V^{(0)}\otimes\phi(\mu)(V^{(0)}_L\otimes\tilde{\phi}(\mu))}{V^{(0)}\otimes\tilde{\phi}(\mu)}\nonumber\\
    &=\frac{\alpha_s C_F}{2\pi}\frac{V^{(0)}\otimes\phi(\mu)(V^{(0)}_L\otimes\phi(\mu))\left(1+\mathcal{O}(\alpha_s)\right)}{V^{(0)}\otimes\phi(\mu)\left(1+\mathcal{O}(\alpha_s)\right)}
    \nonumber\\
    &=\frac{\alpha_s C_F}{2\pi}V^{(0)}_L\otimes\phi(\mu)+\mathcal{O}(\alpha^2_s),
\end{align}
which cancels the $\mu$ dependence in $\mathcal{C}^{(1)}_L(\mu)\otimes\phi(\mu)$. 
So we write the matching formula as
\begin{align}
    \tilde{\phi}(x)=&\frac{V^{(0)}_L\otimes\tilde{\phi}}{V^{(0)}\otimes\tilde{\phi}}\phi(x,\mu_1)+\mathcal{C}^{(1)}_L(\mu_1)\otimes\phi(\mu_1)\nonumber\\+&\frac{V^{(0)}_R\otimes\tilde{\phi}}{V^{(0)}\otimes\tilde{\phi}}\phi(x,\mu_2)+\mathcal{C}^{(1)}_R(\mu_2)\otimes\phi(\mu_2),
\end{align}
where the two scales are chosen to be $\mu_1=2xP_z$ and $\mu_2=2(1-x)P_z$ so that the small-momentum logarithms in both $\mathcal{C}^{(1)}_L(\mu_1)$ and $\mathcal{C}^{(1)}_R(\mu_2)$ vanish.

To extract the lightcone-DA at a specific factorization scale $\mu$, we can relate the quasi-DA, $\tilde{\phi}(x)$, to the lightcone DA, $\phi(\mu,x)$, through
\begin{align}
    \tilde{\phi}&=\left(\left(\frac{V^{(0)}_L\otimes\tilde{\phi}}{V^{(0)}\otimes\tilde{\phi}}+\hat{\mathcal{C}}^{(1)}_L(\mu_1)\right)\mathcal{P}\exp\left[\int_\mu^{\mu_1} \hat{V}d\ln\mu'^2\right]\right.\nonumber\\
    &\left.+\left(\frac{V^{(0)}_R\otimes\tilde{\phi}}{V^{(0)}\otimes\tilde{\phi}}+\hat{\mathcal{C}}^{(1)}_R(\mu_2)\right)\mathcal{P}\exp\left[\int_\mu^{\mu_2} \hat{V}d\ln\mu'^2\right]\right)\phi(\mu)\nonumber\\
    &=\hat{\mathcal{C}}_\text{RGR}(\mu)\phi(\mu),
\end{align}
where $\hat{\mathcal{C}}$ and $\hat{{V}}$ are both operators acting on the function $\phi$ through a convolution.
Once we solve the resummed matching kernel, $\hat{\mathcal{C}}_\text{RGR}(\mu)$, the lightcone DA can be extracted from the inverse matching,
\begin{align}
    \phi(\mu,x)=\int dy \hat{\mathcal{C}}^{-1}_\text{RGR}(x,y,\mu,P_z)\tilde{\phi}(y,P_z).
\end{align}

An easy way to implement the RGR for DA numerically is to write all the operators in the matrix representation. Both the quasi-DA and the lightcone DA are vectors on a grid of $x_i$, $\phi_i=\phi(x_i)$. The matching kernel $\hat{\mathcal{C}}$ and the evolution operator $\hat{V}$ are now matrices on a grid of $x_i$ and $x_j$, $\hat{\mathcal{C}}_{ij}=\mathcal{C}(x_i,x_j)$. Thus the resummed matching in matrix representation is
\begin{align}
\label{eq:matching_resum}
    \tilde{\phi}_i&=\left(\left(\frac{V^L_{ij}\tilde{\phi}_{j}}{V_{ij}\tilde{\phi}_{j}}\delta_{ik}+\hat{\mathcal{C}}^L_{ik}(\mu_1)\right)\exp\left[\int_\mu^{\mu_1} \hat{V}d\ln\mu'^2\right]_{kl}\right.\nonumber\\
    &\left.+\left(\frac{V^R_{ij}\tilde{\phi}_{j}}{V_{ij}\tilde{\phi}_{j}}\delta_{ik}+\hat{\mathcal{C}}^R_{ik}(\mu_2)\right)\exp\left[\int_\mu^{\mu_2} \hat{V}d\ln\mu'^2\right]_{kl}\right)\phi_l(\mu)\nonumber\\
    &=\hat{\mathcal{C}}^\text{RGR}_{ij}(\mu)\phi_j(\mu),
\end{align}
where the resummed kernel $\hat{\mathcal{C}}^\text{RGR}$ now is just an $n\times n$ matrix, which can be calculated row by row.

We have used an approximation of $V\otimes\tilde{\phi}=V\otimes\phi+\mathcal{O}(\alpha_s)$ to construct the ratio $(V^{(0)}_L\otimes\tilde{\phi})/(V^{(0)}\otimes\tilde{\phi})$. However, such an approximation is not applicable to all regions, because there are zero points for $V^{(0)}\otimes\phi$. Near this point $x_0$, the higher order term $\sim\mathcal{O}(\alpha_s)$ cannot be ignored, and the ratio also blows up. The point $x_0$ of course depends on the shape of the DA and quasi-DA curve. A test on several functions suggests that $x_0\sim0.2$ for most DA-like functions. Thus we are still able to perform the above resummation within the region $0.25<x<0.75$. In principle, the formalism may also work for $x<0.1$ and $x>0.9$.

Another issue of constructing the matching matrix is the endpoint regions of $x$. When either the scale $\mu_1=2xP_z$ or $\mu_2=2(1-x)P_z$ becomes very small, $\alpha_s(\mu)$ increases to $\mathcal{O}(1)$. In this region, the perturbative expansion in $\alpha_s(\mu)$ fails, thus the matching cannot be determined through perturbative calculations. In the numerical evaluation, these rows may blow up. Since we are calculating the matching kernel in perturbation theory, the inverse matching kernel at NLO can be expressed as
\begin{align}
    \mathcal{C}^{-1}(x,y,\mu,P_z)=\delta(x-y)-\mathcal{C}^{(1)}(x,y,\mu,P_z)+\mathcal{O}(\alpha_s^2).
\end{align} 
Applying to our matrix form, we get the inverse matching matrix
\begin{align}
    \hat{\mathcal{C}}^{\text{RGR},-1}_{ij}=2\delta_{ij}-\hat{\mathcal{C}}^\text{RGR}_{ij},
\end{align}
which does not need the full information of the matrix to obtain a part of its inverse. So the endpoint region is no longer a problem for us to extract the resummed DA $\phi(x,\mu=2\text{ GeV})$ in the intermediate $x$ region. Those endpoint regions are thus left undetermined by perturbative calculations, and may be obtained from some nonperturbative approaches in the future.

With this approach, we are able to extract the lightcone DA in the intermediate region $0.25<x<0.75$.


\section{Constraints from Short-distance OPE}\label{sec:ope}
LaMET allows us to determine the pion DA in the moderate $x$-region. The calculation begins to break down when $x\to0$ or $x\to1$ as shown in Eq.~\eqref{eq.qDAandDA}. We can, however, determine the global behavior of the pion DA from the short-distance OPE and use this information to constrain the endpoints. 

At short distances, the renormalized coordinate space matrix elements can be expanded in terms of the Mellin moments of the DA
\begin{align}
\label{eq:moment}
    \expval{\xi^n}=\int_0^1dx\phi_{\pi}(x,\mu)(2x-1)^n
\end{align}
where $\xi=x-(1-x)=2x-1$, via the OPE,
\begin{align}
    \tilde{H}^R(z,P_z)=\sum_{n=0}^{\infty}\frac{(-izP_z/2)^n}{n!}\sum_{m=0}^nC_{nm}(z,\mu)\expval{\xi^m}
    +\mathcal{O}(z^2\LambdaQCD^2)
\end{align}
where $C_{nm}(z,\mu)$ are the Wilson coefficients, and $\mathcal{O}(z^2\LambdaQCD^2)$ are higher twist effects which become relevant at distances $z\gtrsim 0.2$~fm. Fitting the matrix elements at short distances to the relevant Wilson coefficients allows us to determine the Mellin moments, $\expval{\xi^n}$. The Mellin moments describe the global information of the DA and, thus, set constraints on the endpoint region once the mid-$x$ distribution is determined. Combined with the physical requirement that $\phi(0)=\phi(1)=0$, it allows us to obtain the shape of the DA near the endpoints with small model dependence. This is called the ``complementarity'' between the large momentum expansion and the short distance factorization. 

In our DA analysis, we can fit the moments from the following RG invariant and renormalization independent ratio~\cite{Gao:2022vyh}:
\begin{align}\label{eq:ratio_ope}
    \mathcal{M}(z,P_1,P_2)=\tilde{H}^B(z,P^z_1)/\tilde{H}^B(z,P^z_2)=\frac{\sum_{n=0}^{\infty}\frac{(-izP_1^z/2)^n}{n!}\sum_{m=0}^nC_{nm}(z,\mu)\expval{ \xi^m}}{\sum_{n=0}^{\infty}\frac{(-izP_2^z/2)^n}{n!}\sum_{m=0}^nC_{nm}(z,\mu)\expval{\xi^m}},
\end{align}
which can be truncated at some order because higher moment contributions are negligible. Then with the Wilson coefficients known from the perturbative calculation, a fit to the short-distance ratio $\mathcal{M}(z,P_1,P_2)$  determines the second Mellin moments $\expval{\xi^2}$ independent of the renormalization method. Given the $\phi^L(x,\mu)$ calculated from LaMET in the mid-$x$ region, we can model the full $x$-dependence, $\phi^f(x,\mu)$, as 
\begin{align}\label{eq.PhiFull}
    \phi^f(x,\mu)=
    \begin{cases}
        \phi^L(x_0,\mu)x^m/x_0^m & 0\leq x\leq x_0\\
        \phi^L(x,\mu)&  x_0\leq x\leq1-x_0\\
        \phi^L(x_0,\mu)\overline{x}^m/\overline{x}_0^m & 1-x_0\leq x\leq1
    \end{cases}
\end{align}
where $x_0$ is the minimal $x$ we can calculate with LaMET. Then we can determine the parameter $m$ by requiring
\begin{align}
    \int_0^1 dx \phi^f(x,\mu)(2x-1)^2=\expval{ \xi^2}_\text{OPE}(\mu)
\end{align}
to obtain the full distribution. 

An alternative approach is not to constrain the endpoint region with moments, but with short distance correlations. Constructing the same full $x$-range distribution $\phi^f(x,\mu)$, we can first Fourier transform it to coordinate space
\begin{align}
    h^f(z,P_z,\mu)=\int_0^1 dx e^{i xzP_z} \phi^f(x,\mu),
\end{align}
then use the short-distance factorization in Eq.~\eqref{eq:sdf} to convert it to the quasi-DA correlations $\tilde{h}^f(z,P_z,\mu)$, and fit to our renormalized matrix elements $\tilde{h}(z,P_z,\mu)$. This approach depends on the renormalization of our matrix elements in coordinate space, but not on the data at other momenta. 

We expect the second approach to give the same result as the first one. The two approaches provide a consistency check for our renormalization method with the short-distance OPE.


\section{Numerical Results}\label{sec:numerical}

In this work, we re-analyze the data presented in Ref.~\cite{Zhang:2020gaj}, measured on three lattice ensembles of lattice spacings $a=\{0.0582(4),0.0888(8),0.1207(11)\}$~fm and pion mass $m_\pi\approx 310$~MeV generated by the MILC collaboration~\cite{MILC:2012znn}. The analysis starts with the same bare matrix elements extracted from a two-state fit to the lattice correlators.


\subsection{Renormalization}\label{sec:Renormalization}
The method used in Ref.~\cite{Zhang:2020gaj} for renormalization was the regularization-independent momentum subtraction (RI/MOM) scheme. However, this method has some problems when dealing with the linear divergence in the nonlocal operator with a spatial Wilson line, as well as generating unknown nonperturbative effects at large distances~\cite{Zhang:2020rsx}. We deal with these issues by working in the hybrid scheme~\cite{Ji:2020brr} with the LRR-improved ratio scheme at short distances from the self-renormalization~\cite{LatticePartonCollaborationLPC:2021xdx}, as discussed in the previous section. The renormalization factors $Z_R(z,a)$ at short distances are obtained from $P_z=0$ matrix elements of the pion PDF to remove the linear divergence,
\begin{equation}
\label{eq:renorm_constant}
    Z_R (z,a,\tau)\equiv\exp\left\{ \Delta\mathcal{I} + \frac{k}{a\ln(a\Lambda_\text{QCD})}-m_0(\tau)z+f(z)a +\frac{3C_F}{4\pi\beta_0}\ln\left[\ln(1/a\Lambda_\text{QCD})\right]+\ln\left[1+\frac{d}{\ln(a\Lambda_\text{QCD})}\right]\right\}
\end{equation}
where $\Delta\mathcal{I}$ is a conversion constant in different schemes, adjusted through fitting to make sure the small $z$-correlations are matched to $\overline{\rm MS}$ results. The fitting parameter $d=-0.53$ represents the NLO RG evolutions on the lattice. The term $\tfrac{k}{a\ln(a\Lambda_\text{QCD})}$ is the linear divergence with fitting parameters $k$ and $\LambdaQCD$ which are not uniquely determined due to the intrinsic ambiguity. By choosing a set of fitting parameters $k=3.3\text{ fm}^{-1}\text{GeV}^{-1}$ and $\Lambda_\text{QCD}=0.1$~GeV~\cite{LatticePartonCollaborationLPC:2021xdx} as scheme $\tau'$, we can determine a corresponding $m_0(\tau)=0.151 \text{ GeV}$~\cite{lrr} to relate the $P_z=0$ lattice matrix elements to the perturbative calculation of $C^\text{LRR}_{00}(z,\mu,\tau)$, as defined in Sec.~\ref{sec:renormalon} with the asymptotic form, which eliminates such an ambiguity. The term $f(z)a$ as a fit parameter incorporates the discretization effects and the remaining terms come from the resummed logarithmically divergent dependence on $a$. 

\begin{figure}[thbp!]
    \centering
    \includegraphics[width=0.5\linewidth]{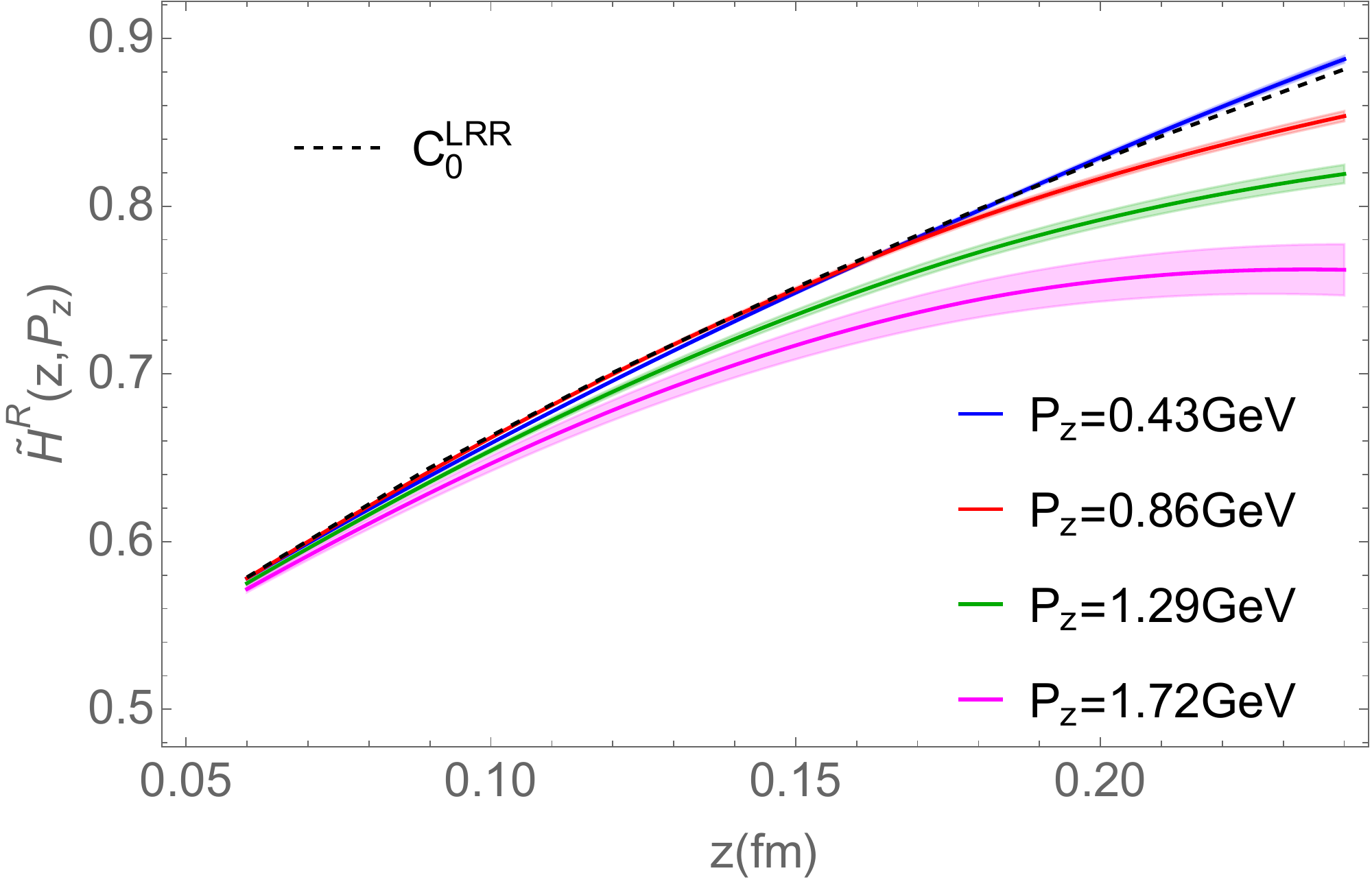}
    \caption{The renormalized matrix elements for different momenta. As the momentum decreases, the distribution approaches the $P_z=0$ perturbative calculations $C^\text{LRR}_0(z,z^{-1})$.}
    \label{fig:renormalization}
\end{figure}

After removing the linear and logarithmic UV divergences through Eq.~\eqref{eq.Ratio}, we are able to extrapolate to the continuum limit to take out additional discretization effects at finite $P_z$, which is a simple process of fitting the renormalized matrix elements at different lattice spacings but at a fixed $z$ value to a linear function:
\begin{equation}
    \tilde{H}^R(z,a,P_z)=c(z)\times a+\tilde{H}^R(z,a=0,P_z).
\end{equation}
for some function $c(z)$ where $\tilde{H}^R$ is defined in Eq.~\eqref{eq:phase_rot}. We carry out this extrapolation for a continuous curve after interpolating our data on three lattice spacings. 
\begin{figure}[thbp!]
    \centering
    \includegraphics[width=0.5\linewidth]{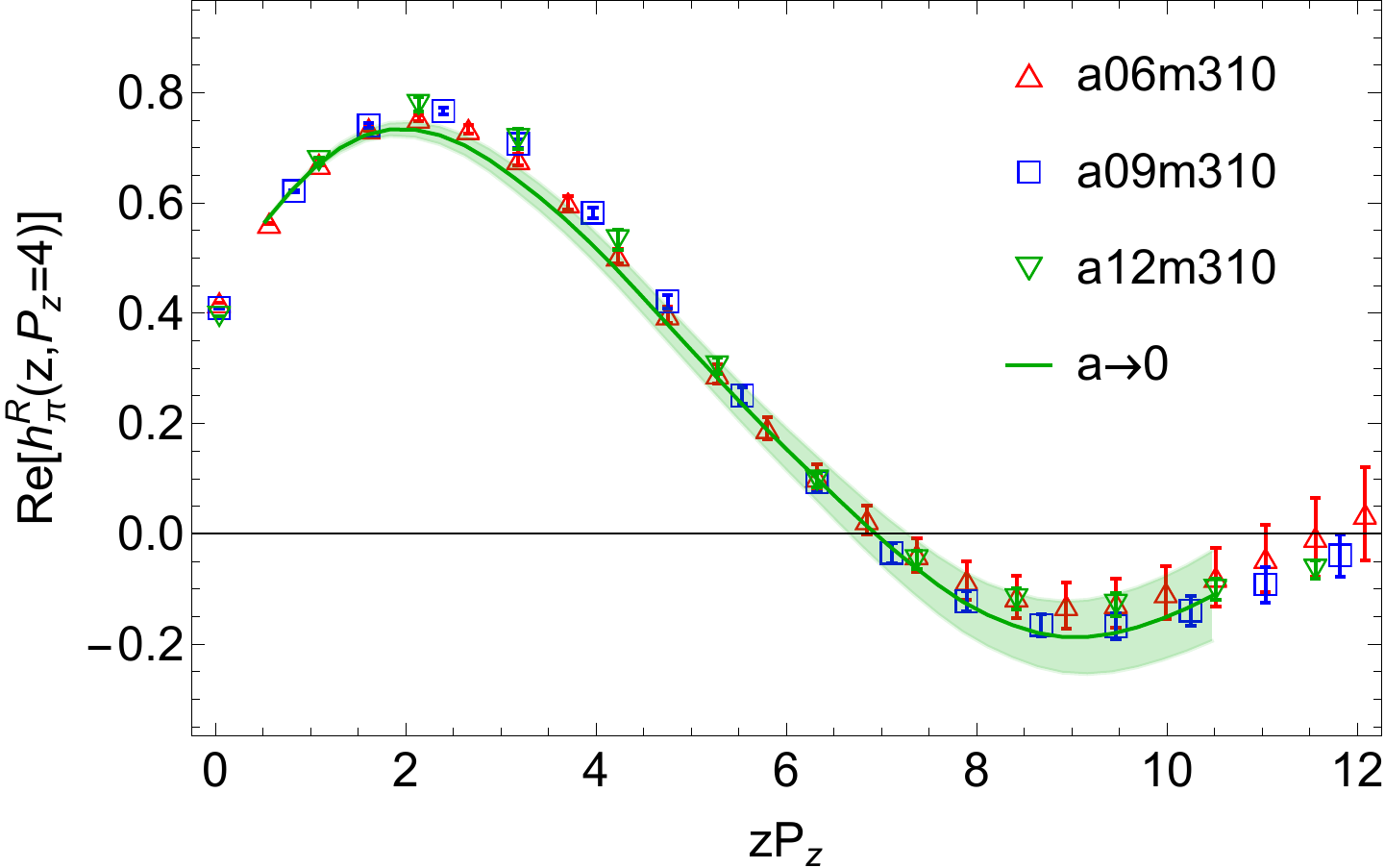}
    \caption{The continuum extrapolation for renormalized phase rotated matrix elements at $P_z=1.72$~GeV. For the pion, the phase rotated matrix elements are purely real. There is a good consistency among different ensembles, so the continuum extrapolation works fine.}
    \label{fig:renormalization_continuum}
\end{figure}
The matrix element after renormalization and continuum extrapolation is shown in Fig.~\ref{fig:renormalization_continuum}, which shows a good consistency among different lattice spacings. A comparison of renormalized matrix elements $\tilde{H}^R$ at different momenta with the LRR perturbative result at $P_z=0$ is shown in Fig.~\ref{fig:renormalization}. We can clearly see that the distribution approaches the $P_z=0$ perturbative calculations $C^\text{LRR}_{00}(z,z^{-1})$ when the momentum decreases.

Then a ratio to the $P_z=0$ perturbative results for DA is taken at short distance $|z|<z_s$ to convert to the hybrid scheme,
\begin{align}
    \mathcal{M}^\text{hybrid}(z,P_z,0)=\frac{\tilde{H}^\text{R}(z,P_z)}{\tilde{H}^R(z,P_z=0)}\theta(z_s-|z|)+\frac{\tilde{H}^\text{R}(z,P_z)}{\tilde{H}^R(z_s,P_z=0)}\theta(|z|-z_s).
\end{align}
where $\tilde{H}^{\rm R}$ is defined in Eq.~\eqref{eq:phase_rot}. The $P_z>0$ matrix elements $\tilde{H}^\text{R}(z,P_z)$ are obtained from our data renormalized by Eq.~\ref{eq:renorm_constant}, while the $\tilde{H}(z,P_z=0)\propto P_z=0$ vanishes in DA measurement, thus we use the Wilson coefficient $C_{00}$ from the perturbative calculation in $\overline{\rm MS}$ scheme. The parameter $z_s$ that separates the nonperturbative and perturbative regions must be much larger than the lattice spacing to avoid discretization effects but not so large as to necessitate higher-twist terms in the OPE. In our calculations, we choose $z_s=0.18$~fm. Note that there is
no modification of $z$ dependence in the second term to avoid introducing unwanted non-perturbative effect.

\begin{figure}[thbp!]
    \centering
    \includegraphics[width=0.5\linewidth]{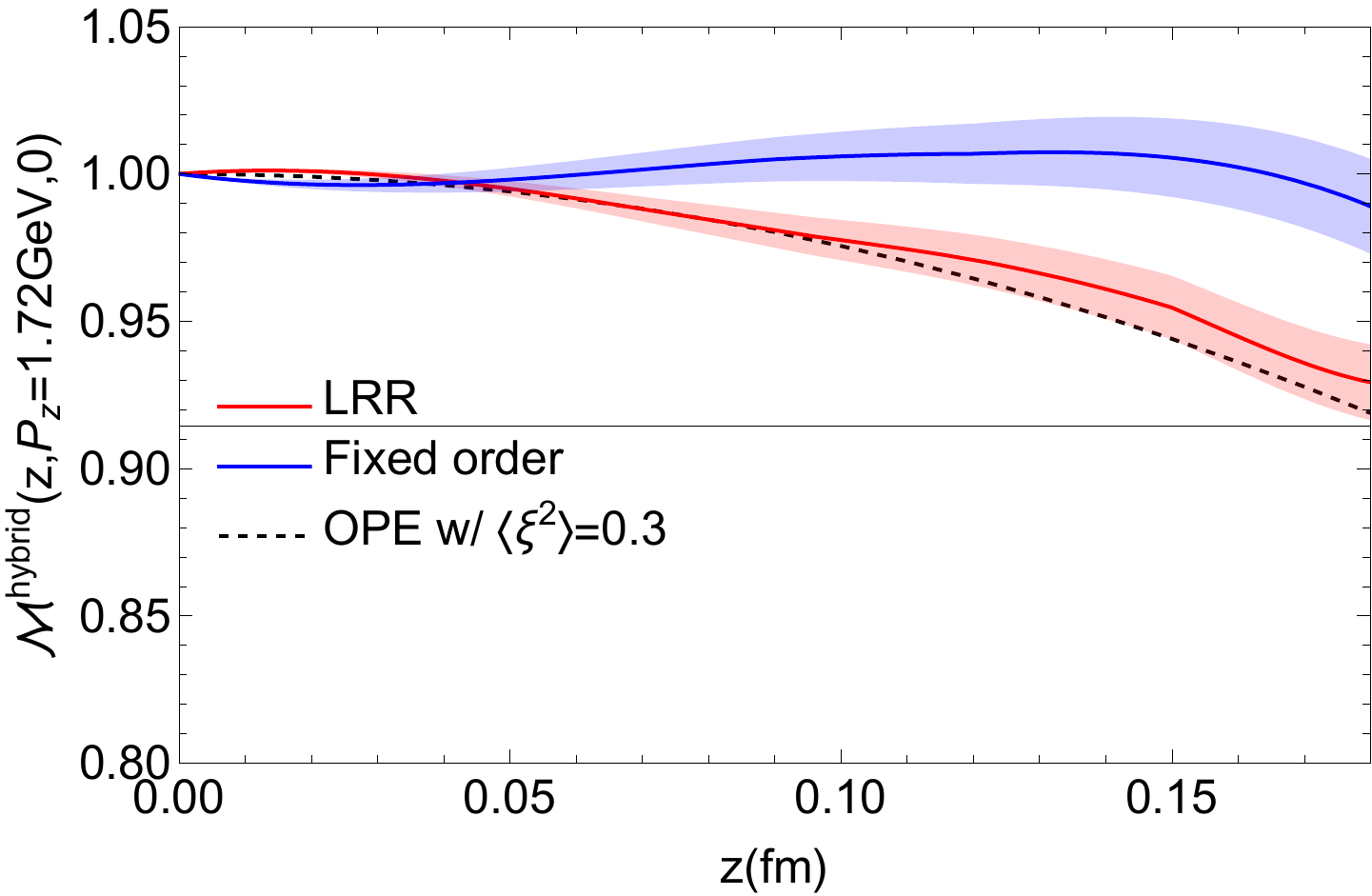}
    \caption{A comparison between the fixed-order self-renormalization and the LRR-improved self-renormalization. The LRR improves the short distance behavior to be more consistent with the OPE predictions.}
    \label{fig:lrr_renorm_eff}
\end{figure}

Figure~\ref{fig:lrr_renorm_eff} shows a comparison between the ratio $\mathcal{M}^\text{hybrid}(z,P_z,0)$ obtained from fixed-order self-renormalization and the LRR-improved self-renormalization. An obvious problem in the fixed-order renormalization is the hump near $z=0.1$~fm, which suggests a negative second moment $\expval{\xi^2}<0$, irreconcilable with the OPE predictions at short distance. The LRR-improved renormalized matrix elements, on the other hand, show good consistency with OPE predictions at short distances. This comparison demonstrates that the modification from LRR is necessary for a correct renormalization.  Both the continuum extrapolation and the conversion to the hybrid scheme are linear, thus the two steps commute with each other.


\subsection{Extracting $x$-dependence of quasi-DA}
In order to Fourier transform our coordinate space correlations to momentum space and extract the $x$-dependence, we need to extrapolate our matrix elements to infinite distance. We first convert our position-space variable to quasi light-cone distance $\lambda\equiv zP_z$ as introduced in Sec.~\ref{sec:resummation}. Since $P_z$ is fixed for a single calculation, large distance corresponds to large $\lambda$. Although the large-$\lambda$ correlation becomes extremely noisy at a finite momentum from the lattice, which in principle makes it impossible for us to know the longtail information, the distribution is not arbitrary. A general consideraction of
coordinate space correlations suggests an algebraic decay along with an exponential decay~\cite{Ji:2020brr,LatticeParton:2022zqc}. The constraints allow us to reduce the error in the large-$\lambda$ region and extract the $x$-dependence of the quasi-DA.

\begin{figure}[thbp!]
    \centering
    \includegraphics[width=0.5\linewidth]{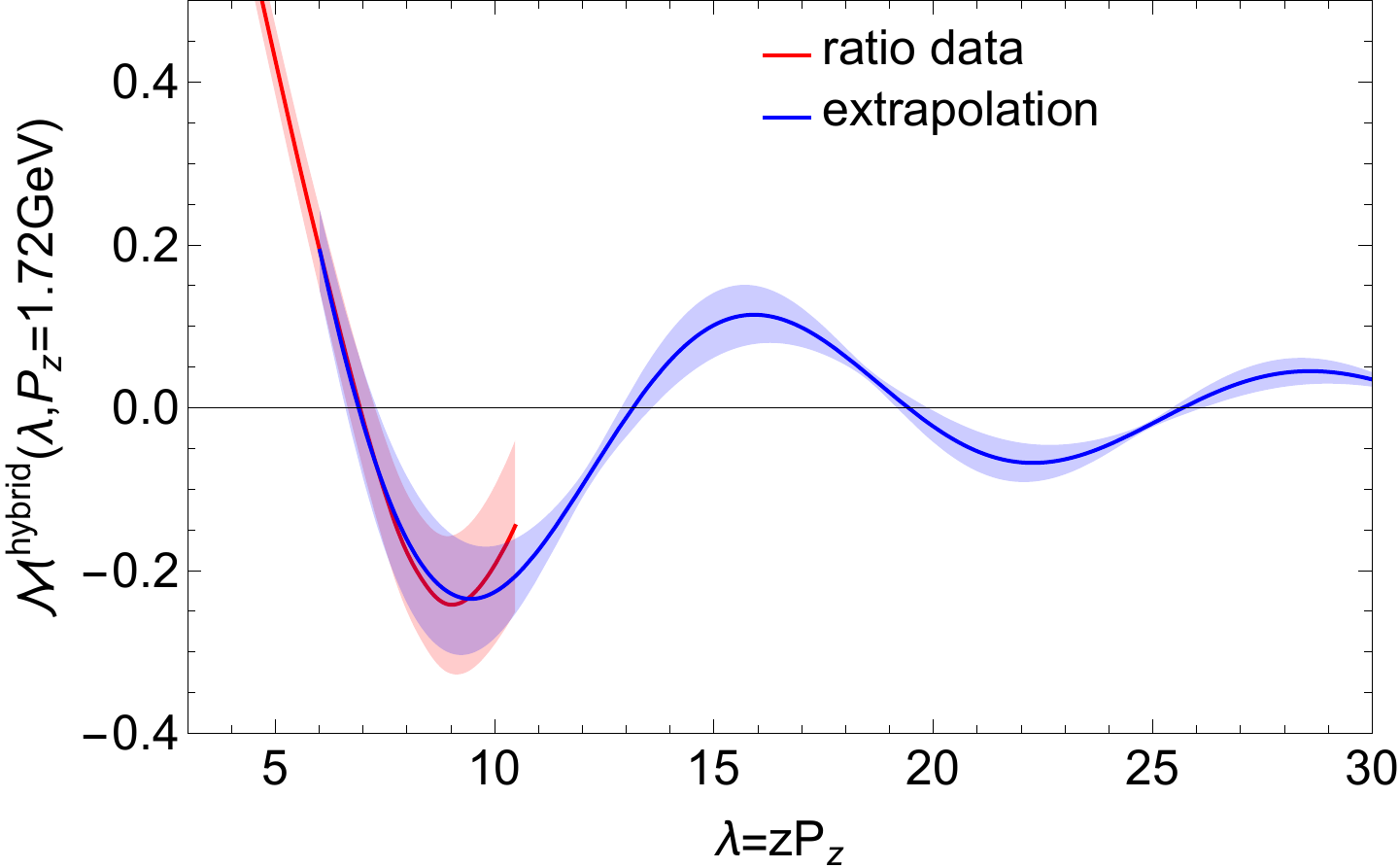}
    \caption{Continuum position space ratio in the hybrid scheme, $\mathcal{M}(\lambda)$, with extrapolations to large-$|\lambda|$. The consistency between the data (red) and the extrapolation (blue) in the overlapping region suggests the good quality of the extrapolation.}
    \label{fig:largelambdaextrpn}
\end{figure}

Based on these constraints, we can extrapolate our matrix elements in position space to the corresponding inverse Fourier transform~\cite{Ji:2020brr,LatticeParton:2022zqc}:
\begin{equation}\label{eq.largelambda}
    \tilde{H}^\text{R}(\lambda\to\infty,P_z)=\left(e^{i\lambda/2}\frac{c_1}{(i\lambda)^n}+e^{-i\lambda/2}\frac{c_1}{(-i\lambda)^n}\right)e^{-|\lambda|/\lambda_0}
\end{equation}
where $\lambda_0$ is a large constant describing the correlation length and depends on the hadron momentum, and the terms $(c_1,n)$ are fitting parameters. Note that at long distances in our hybrid scheme, the ratio $\mathcal{M}^{\rm hybrid}(\lambda,P_z)$ only differs from $\tilde{H}^\text{R}(\lambda,P_z)$ by a constant factor $\tilde{H}^\text{R}(z_s,0)$, so they have the same functional form. We then fit the longtail of $\mathcal{M}^{\rm hybrid}(\lambda,P_z)$ to Eq.~\eqref{eq.largelambda}. An example of the large-$\lambda$ extrapolation is shown in Fig.~\ref{fig:largelambdaextrpn}.

\begin{figure}[thbp!]
   \centering
   \includegraphics[width=0.5\linewidth]{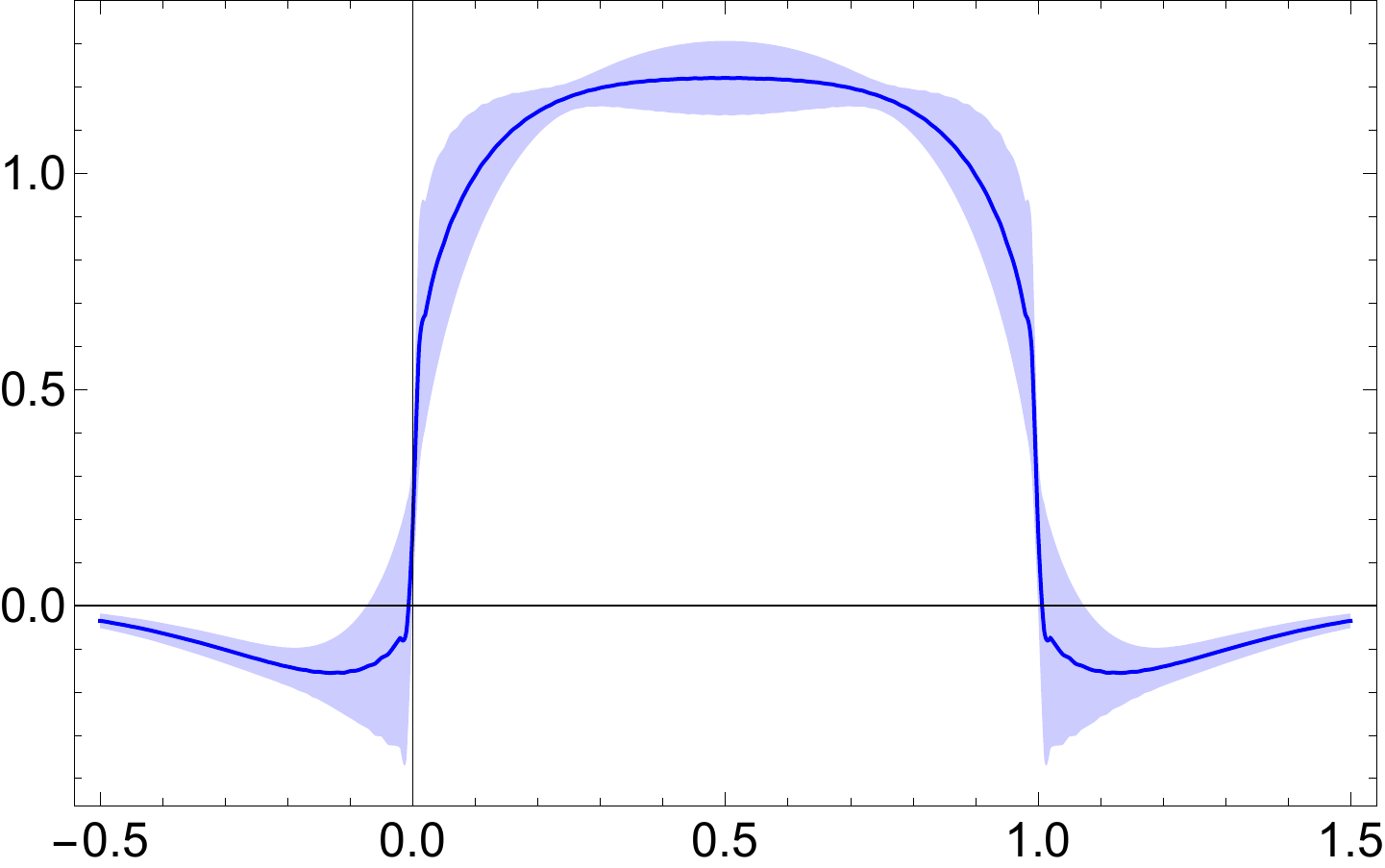}
   \caption{The $x$-dependence of the quasi-DA. There is a nonvanishing distribution outside the physical region $x\in[0,1]$ for the quasi-DA. This nonphysical distribution will be suppressed when matched to the lightcone.}
   \label{fig:PionqDA_x}
\end{figure}

\begin{figure}[thbp!]
   \centering
   \includegraphics[width=0.5\linewidth]{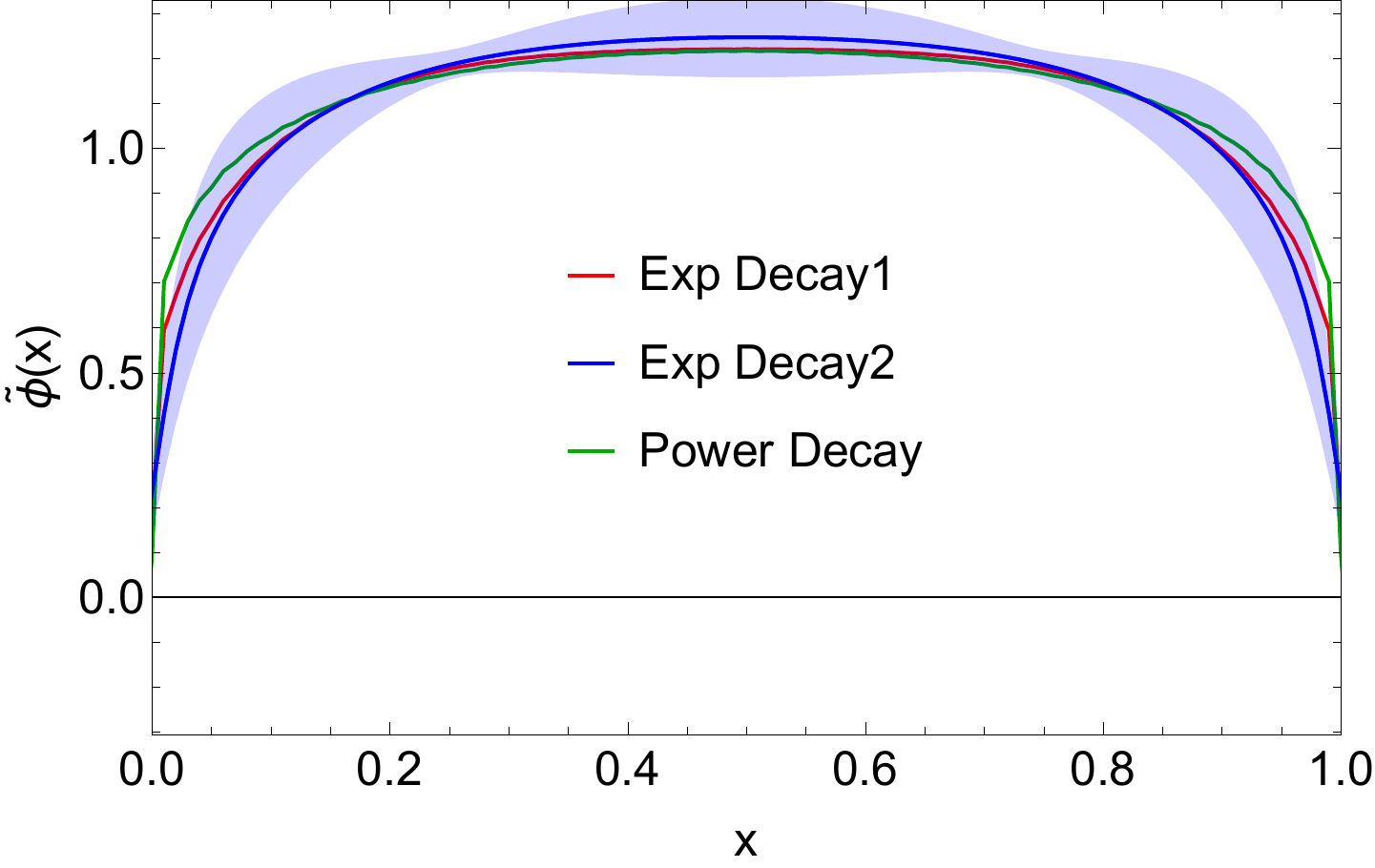}
   \caption{The $x$-dependence of the quasi-DA from three different long-tail modelings. Their results are consistent, with a difference much smaller than the statistical error (blue band) in the mid-$x$ region.}
   \label{fig:PionqDA_x_model}
\end{figure}

With a full-range coordinate space correlation, we are able to extract $\tilde{\phi}(x,P_z)$ through a Fourier transformation in our chosen renormalization scheme,
\begin{equation}\label{eq.FourierTransform}
    \tilde{\phi}_{\pi}(x,P_z)=\int^{\infty}_{-\infty}\frac{d\lambda}{2\pi}e^{ix\lambda}\mathcal{M}^\text{hybrid}(\lambda,P_z)
\end{equation}
as shown in Fig.~\ref{fig:PionqDA_x}.
Although we used a model to describe the large-$\lambda$ behavior, we should address that the final result is not sensitive to the highly suppressed long tails. To illustrate this, we use two different models corresponding to Eq.~\eqref{eq.largelambda}, one is not to include the exponential decaying factor (labelled as ``Power Decay''), the other is by fixing the correlation length $\lambda_0=50$ (labelled as ``Exp Decay2'') to check the sensitivity of DA to the long-tail model assumptions. The comparison is shown in Fig.~\ref{fig:PionqDA_x_model}, showing small discrepancies in mid-$x$ region when compared to the statistical error in blue band. The endpoint regions are more sensitive to the long-tail modeling, but we only calculate the mid-$x$ region of light-cone DA directly and the endpoint regions are obtained from modeling. So the long-tail modeling dependence has little influence on our final determination of the full-x distribution.

\subsection{Matching to obtain DA in mid-$x$ region}

\begin{figure}[thbp!]
    \centering
    \includegraphics[width=0.5\linewidth]{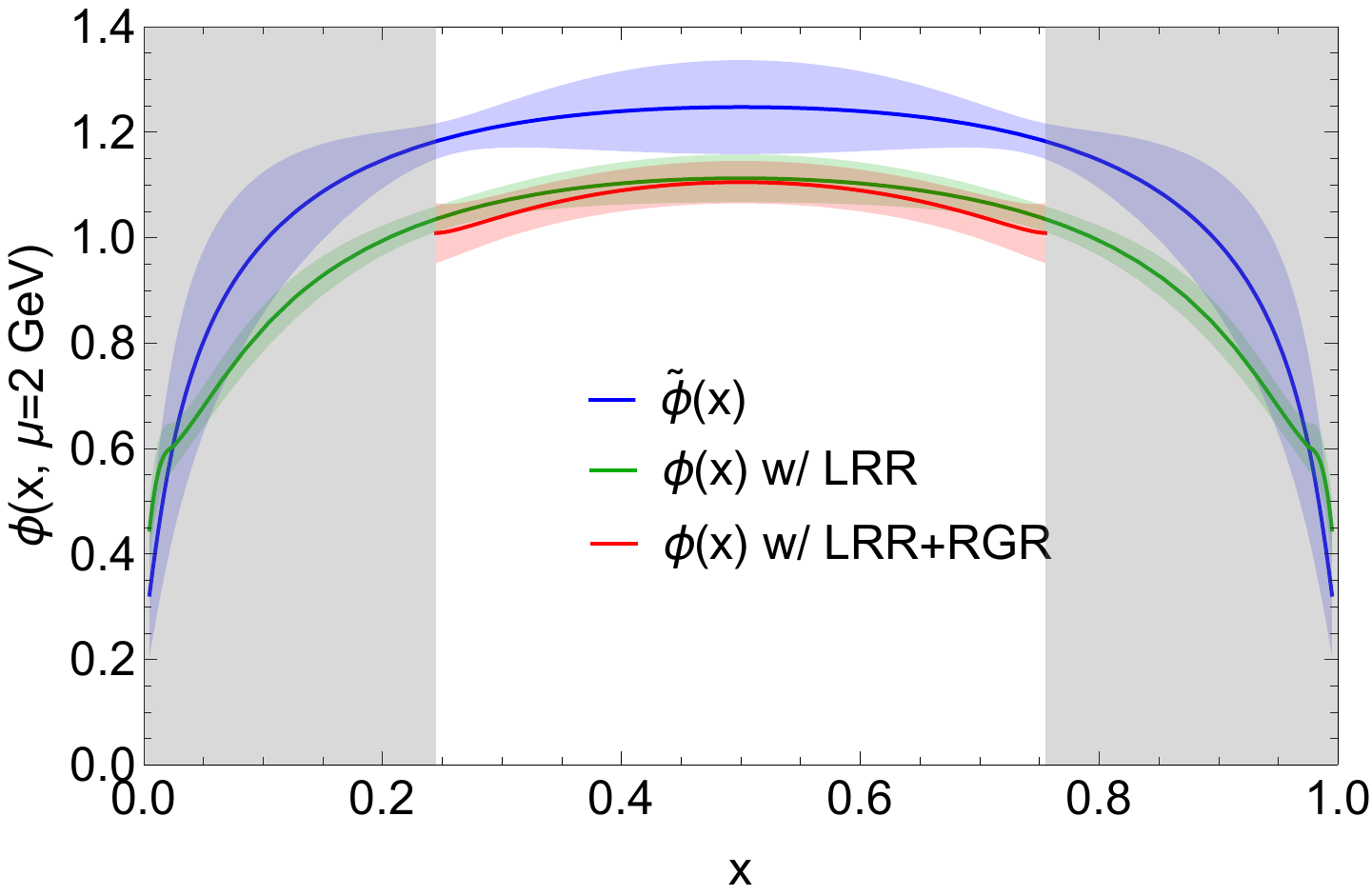}
    \caption{The mid-$x$ dependence of the lightcone DA after we perform the inverse matching on the quasi-DA $\phi(x)$. The blue band is the quasi-DA $\tilde{\phi}(x)$ before matching as a reference. We show only $x\in[0.25,0.75]$ for the result with RGR because the strategy does not work outside the range.}
    \label{fig:mid_x}
\end{figure}

After obtaining the quasi-DA in momentum space, we can then apply the matching to obtain the lightcone DA. Firstly, we apply the fixed-order matching kernel, modified with LRR, at scale $\mu=2$~GeV without the large log resummation. The fixed order matching appears to be valid for the full $x$ region. However, as we discussed, it is just an artificial effect. As we approach the endpoints, the higher-order large logs can no longer be neglected and have to be resummed. We thus apply the matching with RGR, and show the comparison in Fig.~\ref{fig:mid_x}. As we discussed in Sec.~\ref{sec:rgr_th}, the resummed matching causes a divergence at small physical scales, suggesting that the endpoint region is not accessible in perturbation theory, and $x\in[0.25,0.75]$ is considered a safe range for the RGR. So we only show a segment of $x\in[0.25,0.75]$ for the RGR matched result, and use a gray band to shade the outside regions. The RGR effect is almost zero near $x=0.5$, and starts to suppress the distribution when approaching the endpoints. 

\begin{figure}[thbp!]
    \centering
    \includegraphics[width=0.45\linewidth]{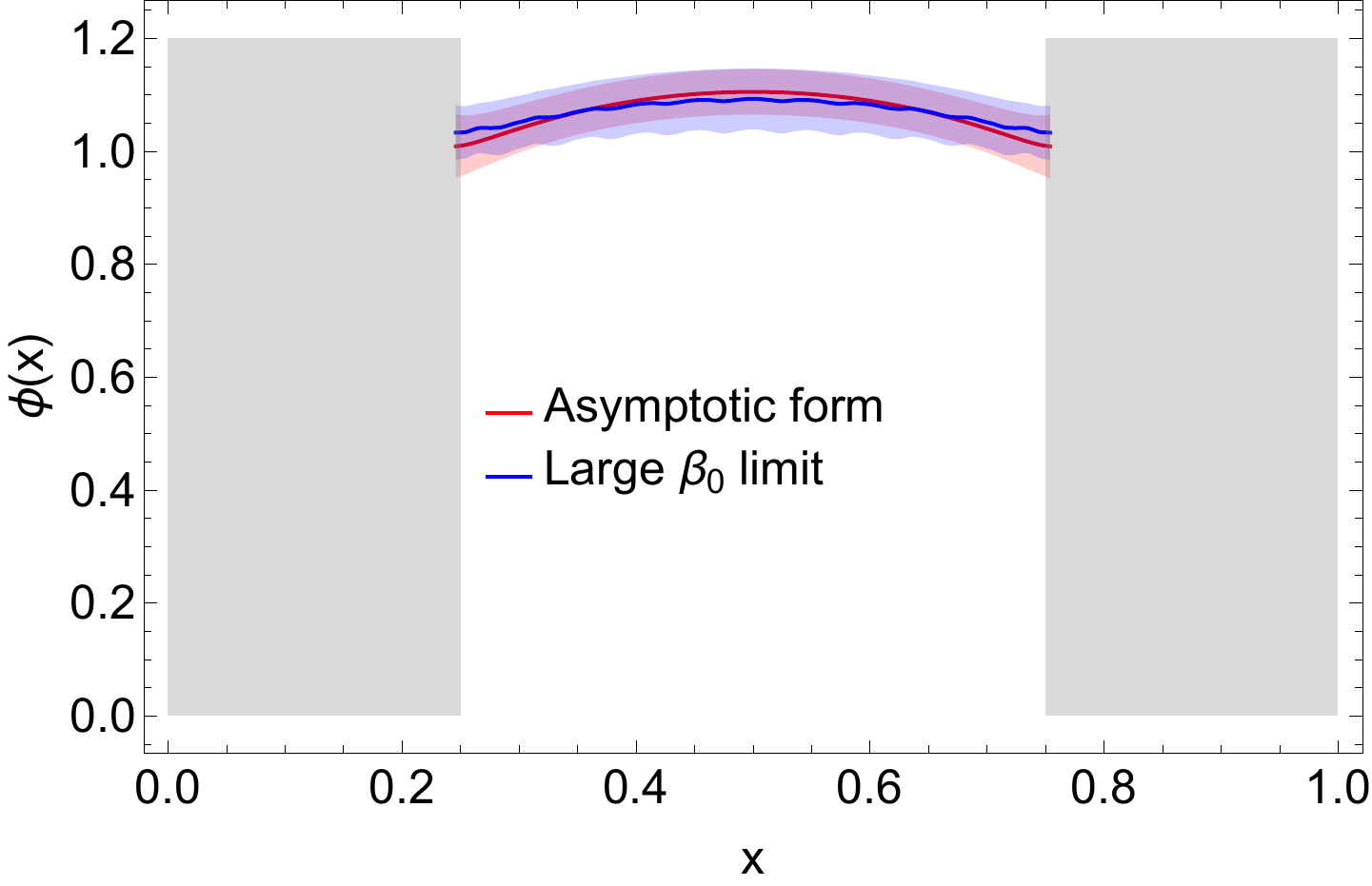}
    \includegraphics[width=0.45\linewidth]{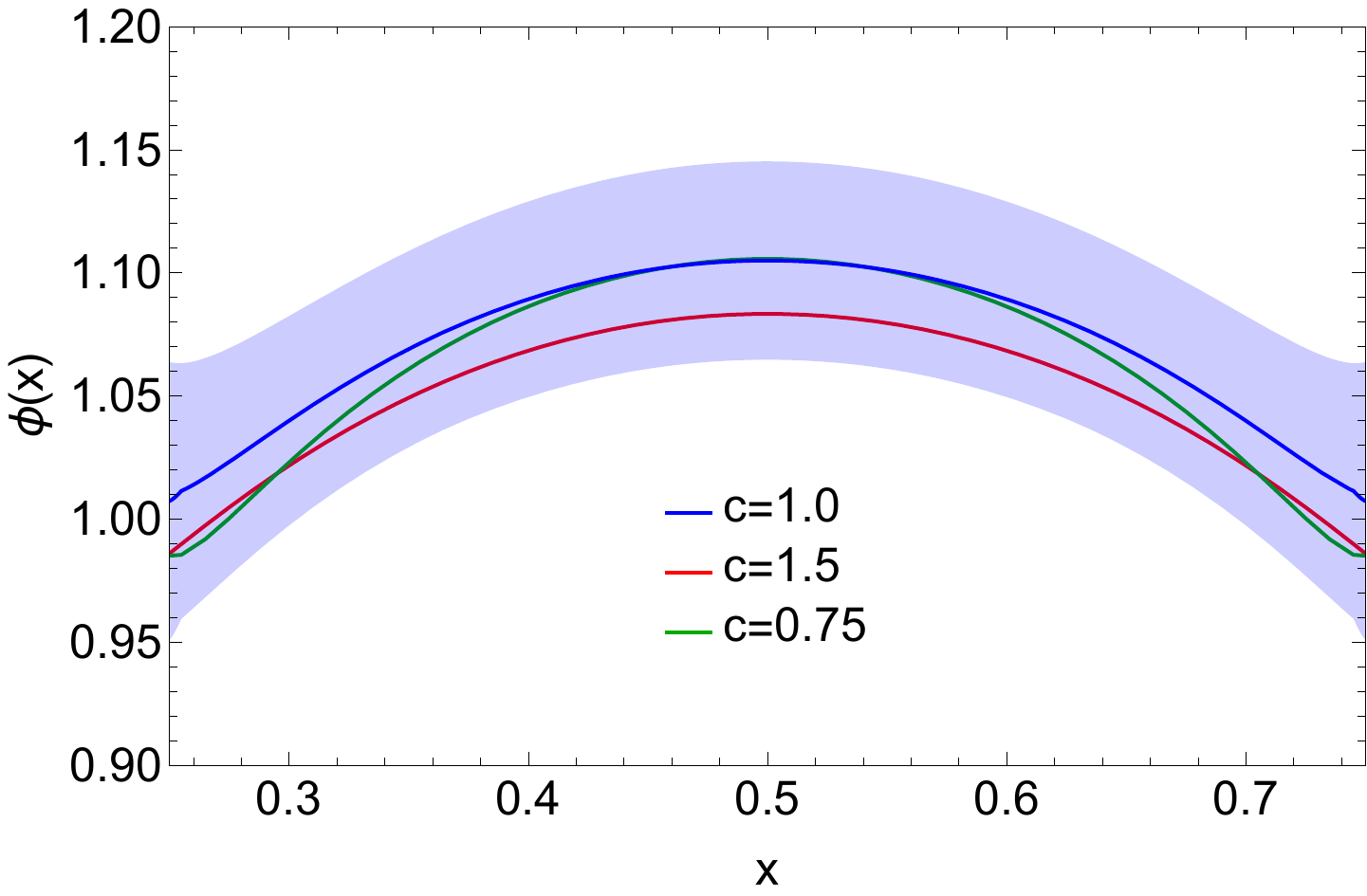}
    \caption{Comparison of the DA in mid-$x$ region after LRR+RGR matching with different LRR approaches (left) and different scale choices (right). They are all consistent within error, up to $<3\%$ difference.}
    \label{fig:sensitivity}
\end{figure}

We also find that the result is insensitive to which LRR method is used, and the scale choice for RGR by changing the initial scale of RGR from $2xP_z$ to $2cxP_z$ with $c\in[0.75,1.5]$, suggesting only $<3\%$ difference, as shown in Fig.~\ref{fig:sensitivity}.

\subsection{Full $x$-dependence for lightcone DA}
\label{sec:rgr_data}

Now we have the distribution determined for the mid-$x$ region, while the endpoint regions are still unknown from the LaMET approach. Fortunately, in coordinate space, the matching coefficients in small-$z$ region is perturbative, thus can be resummed safely. Applying these matching coefficients to lattice data, we are able to obtain the lightcone correlation in a certain range of correlation length $\lambda=zP_z$, which contains the global information of the $x$-dependent DA, such as its moments. With the information from the mid-$x$ region, we can complete our picture of the extracted DA by utilizing the small-$z$ information to constrain the endpoint behavior, as suggested in Ref.~\cite{Ji:2022ezo}.

\begin{figure}[thbp!]
    \centering
    \includegraphics[width=0.5\linewidth]{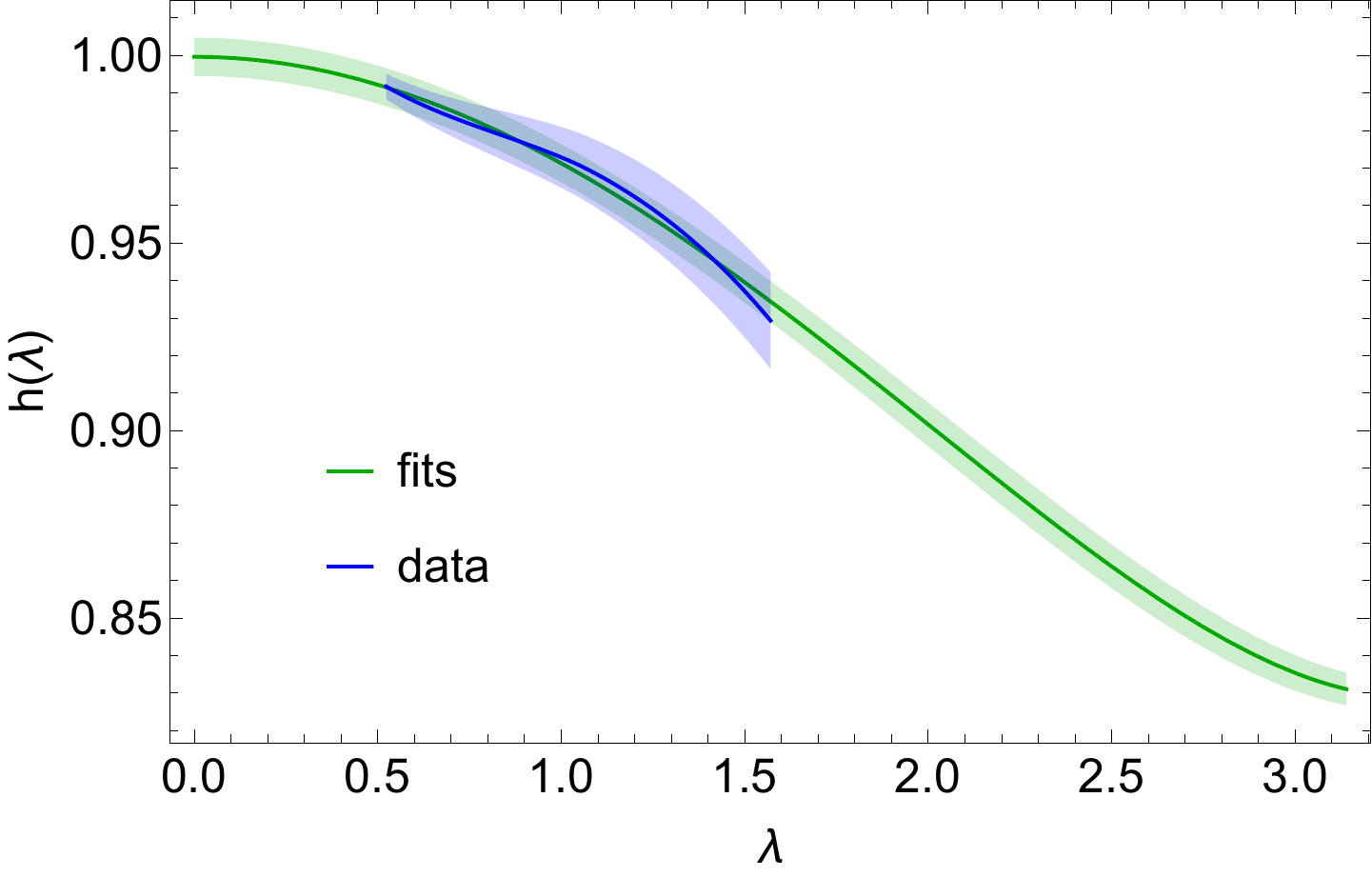}
    \caption{The fitting of parametrized DA to lattice data at short distance. The fitted short-distance correlations agree well with the data.}
    \label{fig:sdf_fit}
\end{figure}

\begin{figure}[thbp!]
    \centering
    \includegraphics[width=0.5\linewidth]{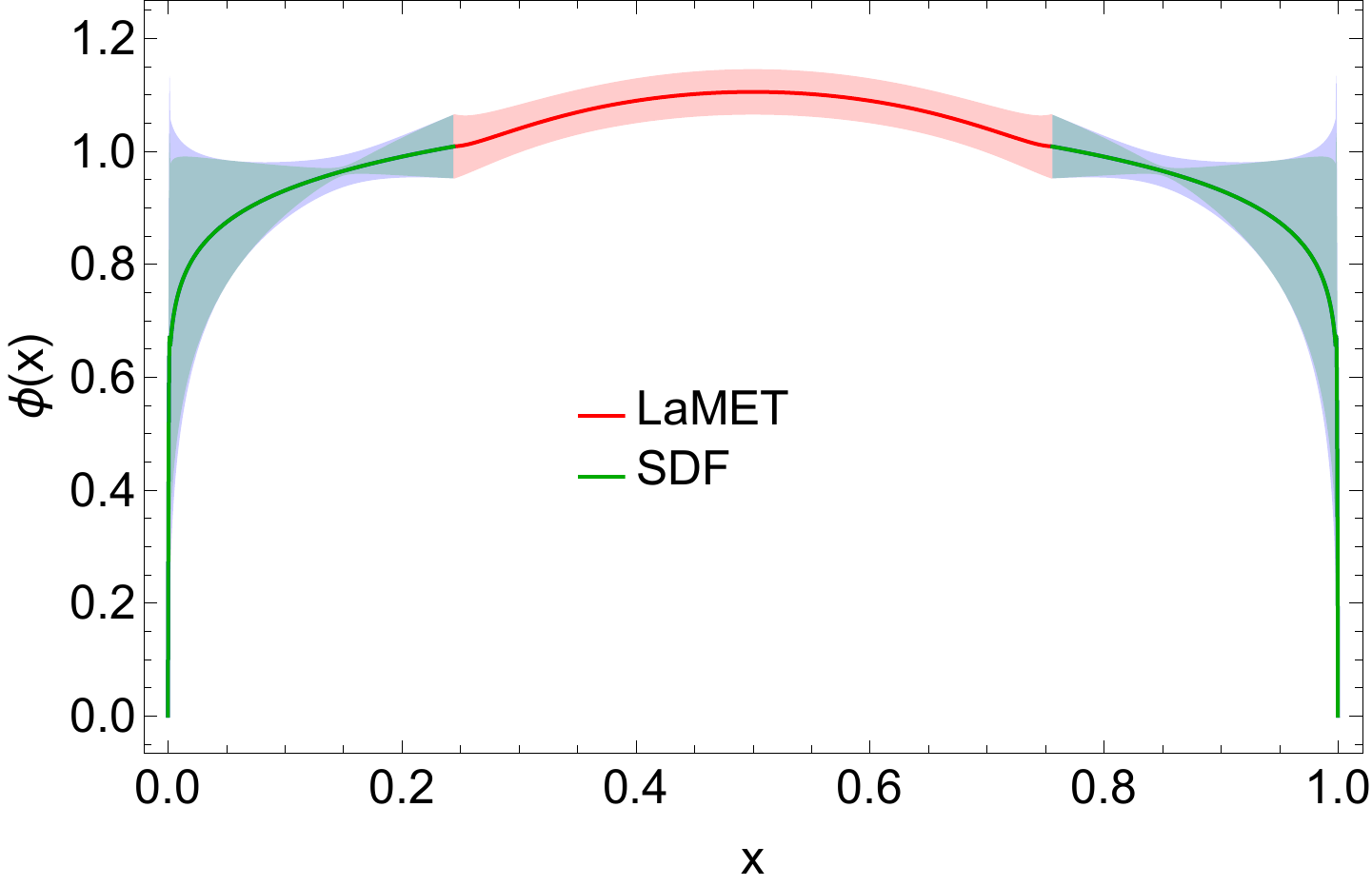}
    \caption{Full $x$ dependence of the DA. The red band in $x\in[0.25,0.75]$ is obtained from LaMET, and the green (blue) band is obtained from the short-distance correlations by modeling of the endpoints with statistical (statistical and systematic) errors.}
    \label{fig:full_x_da}
\end{figure}

Near the endpoints, we can parametrize the DA as a power of $x$ or $1-x$ as in Eq.~\eqref{eq.PhiFull}. To ensure continuity, we require that the parametrized form coincide with our mid-$x$ results at $x=x_0$. We convert this parametrized DA into coordinate space, apply the short-distance matching, and fit the result to our renormalized matrix elements. Figure~\ref{fig:sdf_fit} shows the comparison from the parametrized  DA and our lattice data at short distances, which suggests good consistency.
Besides that, we allow some model-dependence by adding a small correction,
\begin{align}
    \phi(x<x_0)=Ax^m(1+\sin(b)x),
\end{align}
where $\sin(b)$ is to guarantee that the size of this correction term is not so large as to cause a sharp turn at the junction point, i.e., the different regions are smoothly connected. The same modification is symmetrically applied to $x\to1-x$, and a model-dependence is included as the systematic error by calculating the difference between the modified model and the original one in Eq.~\eqref{eq.PhiFull}. The final estimation taking into account such a systematic error is shown in Fig.~\ref{fig:full_x_da}.

We can estimate the moment from the full $x$-dependence from Eq.~\eqref{eq:moment} with Eq.~\eqref{eq.PhiFull}. We get
\begin{align}
    \expval{1}=&0.999(5),\quad \expval{\xi^2}=0.306(19),
\end{align}
which are in good agreement with the theoretical normalization $\expval{1}=1$, and the second moment $\expval{\xi^2}=0.298(39)$ obtained from the renormalization-independent OPE fit to Eq.~\eqref{eq:ratio_ope}. This self-consistency is a strong support for our renormalization method.

\begin{figure}[thbp!]
    \centering
    \includegraphics[width=0.5\linewidth]{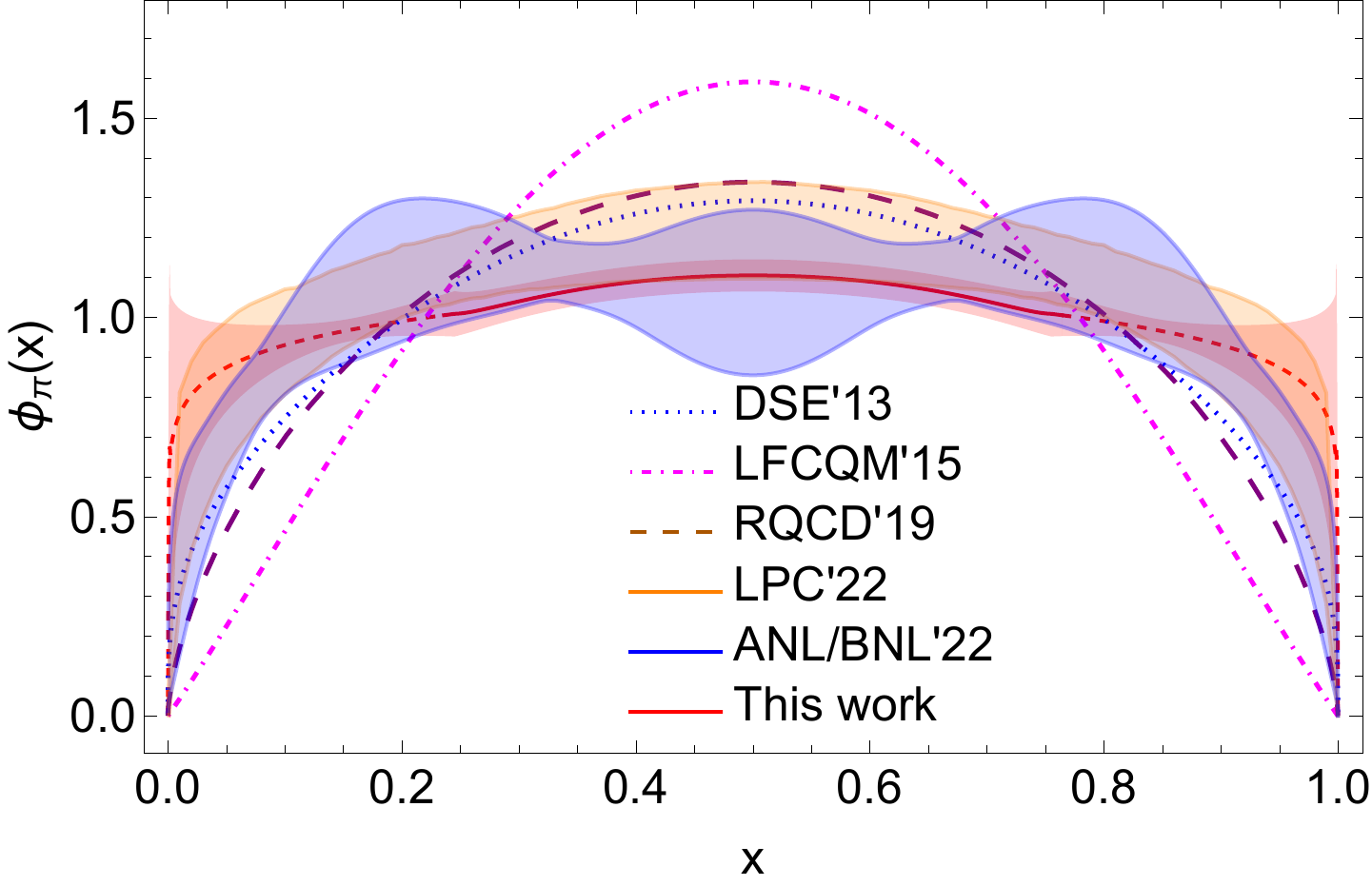}
    \caption{Comparison of the $x$ dependence with previous phenomenological and lattice results. In this work, the mid-$x$ region from direct LaMET calculation is labeled with solid red line, while the endpoint regions from complementarity are labeled with dashed red lines.}
    \label{fig:full_x_comparison}
\end{figure}

In Fig.~\ref{fig:full_x_comparison} we compare with previous model-dependent calculations and lattice results, including the Dyson-Schwinger Equation (DSE'13)~\cite{Chang:2013pq}, the prediction of the light-front constituent-quark model (LFCQM’15)~\cite{deMelo:2015yxk}, the OPE reconstruction from local second moment calculations (RQCD’19)~\cite{RQCD:2019osh}, the lattice calculation from LPC (LPC'22)~\cite{LatticeParton:2022zqc}, and the reconstruction from fitted moments by ANL/BNL collaboration (ANL/BNL'22)~\cite{Gao:2022vyh}.
Our final result suggests the flattest and broadest distribution among all these calculations, as we can also tell from the large second moment in our data. This may have a big impact on the phenomenology of pion hard exclusive processes. At large $Q^2$, the $\pi\gamma\gamma^*$ transition form factor $F_{\pi\gamma\gamma}(Q^2,0)$~\cite{CELLO:1990klc,CLEO:1997fho,BaBar:2009rrj,Belle:2012wwz} and the pion electromagnetic form factor $F_{\pi}(Q^2)$~\cite{JeffersonLab:2008jve} are both sensitive to the shape of the DA $\phi(x,Q^2)$. In general, since both are enhanced near the threshold $x\to 1$~\cite{Melic:1998qr,Gao:2021iqq}, a broader DA will predict the form factors to be larger at large $Q^2$. However, the factorization of the exclusive processes are known to be problematic near the threshold~\cite{Efremov:1979qk,Isgur:1988iw,Li:1992nu}, and the scale setting in the pion electromagnetic form factor $F_{\pi}(Q^2)$ also causes large uncertainty to its estimation from the DA~\cite{Brodsky:1982gc,Melic:1998qr}. Due to these complications, the study of these phenomenologies are beyond the scope of this work, but a more detailed and systematic study is needed in the future to completely understand the impacts.


\section{Conclusion}\label{sec:conclusion}
In this paper we have computed the pion distribution amplitude with momentum fraction in the range $x\in[0,1]$ with improved handling of three sources of systematic errors: removing the $\mathcal{O}(\LambdaQCD/xP_z)$ power correction from intrinsic ambiguities, resumming the small-momentum logarithms, and constraining the distribution near the endpoints from short distance correlations. We renormalize the matrix elements in the hybrid scheme with the LRR-improved self-renormalization factors at short distance. Then an LRR-improved matching kernel is used, along with a two-scale resummation, to obtain the pion DA in the mid-$x$ region $x\in[0.25,0.75]$. We then model the endpoint region with a power law function, allowing a small variation, to reconstruct coordinate space correlations and fit to our data. The second Mellin moment determined from the full-$x$ dependence was $\expval{\xi^2}=0.302(23)$ and from the renormalization-independent short-distance OPE was $\expval{\xi^2}=0.298(39)$. These two results are in good agreement and give us confidence in the determination of the endpoint region of the DA. Our final result suggests a broad distribution of the pion DA. It has the potential for a big impact on the form factors of the DA at large $Q^2$, and will be investigated in detail in the future. 


\subsection*{Acknowledgements}
We thank the MILC Collaboration for sharing the lattices used to perform this study. The LQCD calculations were performed using the Chroma software suite~\cite{Edwards:2004sx}.
This research used resources of the National Energy Research Scientific Computing Center, a DOE Office of Science User Facility supported by the Office of Science of the U.S. Department of Energy under Contract No. DE-AC02-05CH11231 through ERCAP;
facilities of the USQCD Collaboration are funded by the Office of Science of the U.S. Department of Energy,
and the Extreme Science and Engineering Discovery Environment (XSEDE), which was supported by National Science Foundation Grant No. PHY-1548562. 
This research is supported by the U.S. Department of Energy, Office of Science, Office of Nuclear Physics, under contract number DE-SC0020682. J.H. is partially supported by the Center for Frontier Nuclear Science at Stony Brook University. Y.S. is partially supported by the U.S.~Department of Energy, Office of Science, Office of Nuclear Physics, contract no.~DE-AC02-06CH11357. 
The work of HL is partially supported by the US National Science Foundation under grant PHY 1653405 ``CAREER: Constraining Parton Distribution Functions for New-Physics Searches'', grant PHY 2209424,  and by the  Research  Corporation  for  Science  Advancement through the Cottrell Scholar Award.


\appendix
\section{Notations}\label{sec:Notations}
We tabulate the various symbols used throughout this paper for the convenience of the reader in Tab.~\ref{tab:Notation}.
\begin{table}[h]
    \centering
    \begin{tabular}{|c|c|}
    \hline
    Symbol & Definition\\
    \hline\hline
    \multirow{2}{*}{$\tilde{h}^\text{B}(z,P_z,a)$} & \multirow{2}{*}{Bare qDA in coordinate space} \\ & \\
    \hline
    \multirow{2}{*}{$\tilde{h}^\text{R}(z,P_z)$} & \multirow{2}{*}{Renormalized qDA in coordinate space}\\ & \\
    \hline
    \multirow{2}{*}{$\tilde{H}^\text{R,B}(z,P_z)$} & \multirow{2}{*}{$e^{izP_z/2}\tilde{h}^\text{R,B}(z,P_z)$}\\ & \\
    \hline
    \multirow{2}{*}{$\mathcal{M}(z,P^{(1)}_z,P^{(2)}_z)$} & \multirow{2}{*}{$\tilde{H}^\text{B}(z,P_z^{(1)})/\tilde{H}^\text{B}(z,P_z^{(2)})$}\\ &\\
    \hline
    \multirow{2}{*}{$\tilde{\phi}(x,P_z)$} & \multirow{2}{*}{Renormalized qDA in momentum space}\\ & \\
    \hline
    \multirow{2}{*}{$\phi(x,\mu)$} & \multirow{2}{*}{Lightcone DA in momentum space}\\ & \\
    \hline
    \multirow{2}{*}{$\mathcal{C}(x,y,\mu,P_z)$} & \multirow{2}{*}{Momentum space matching kernel}\\ & \\
    \hline
    \multirow{2}{*}{$\mathcal{Z}(\nu,z^2,\mu^2,\lambda)$} & \multirow{2}{*}{Coordinate space matching kernel}\\
    & \\
    \hline
    \multirow{2}{*}{$C_{nm}(z,\mu)$} & \multirow{2}{*}{DA Wilson coefficients}\\ & \\
    \hline
    \end{tabular}
    \caption{Notations used throughout this paper.}
    \label{tab:Notation}
\end{table}

\providecommand{\href}[2]{#2}\begingroup\raggedright\endgroup


\end{document}